\documentclass[aps,prc,floatfix,twocolumn,superscriptaddress,groupedaddress,showpacs]{revtex4-1}
\usepackage{epsfig}
\usepackage{amsmath,fourier}

\begin{document}
\title{Nuclear Dipole Response in the Finite-Temperature Relativistic Time Blocking Approximation}



\author{Herlik Wibowo}
\affiliation{Department of Physics, Western Michigan University, Kalamazoo, Michigan 49008, USA}
\author{Elena Litvinova}
\affiliation{Department of Physics, Western Michigan University, Kalamazoo, Michigan 49008, USA}
\affiliation{National Superconducting Cyclotron Laboratory, Michigan State University, East Lansing, Michigan 48824, USA}


\date{\today}


\begin{abstract}
\noindent\textbf{Background:} 
The radiative neutron capture reaction rates of the r-process nucleosynthesis are immensely affected by the microscopic structure of the low-energy spectra of compound nuclei. The relativistic (quasiparticle) time blocking approximation (R(Q)TBA) has successfully provided a good description of the low-energy strength, in particular, the strength associated with pygmy dipole resonance, describing transitions from and to the nuclear ground state. The finite-temperature generalization of this method is designed for thermally excited compound nuclei and has the potential to enrich the fine structure of the dipole strength, especially in the low-energy region. 

\noindent\textbf{Purpose:} To formulate the thermal extension of RTBA, i.e., finite-temperature relativistic time blocking approximation (FT-RTBA) for the nuclear response, and to implement it numerically for calculations of the dipole strength in medium-light and medium-heavy nuclei.

\noindent\textbf{Methods:} The FT-RTBA equations are derived using the Matsubara Green's function formalism. We show that with the help of a temperature-dependent projection operator on the subspace of the imaginary time it is possible to reduce the Bethe-Salpeter equation for the nuclear response function to a single frequency variable equation also at finite temperatures. The approach is implemented self-consistently in the framework of quantum hadrodynamics and keeps the ability of connecting the high-energy scale of heavy mesons and the low-energy domain of nuclear medium polarization effects in a parameter-free way. 


\noindent\textbf{Results:} The method was applied to the medium-light $^{48}$Ca, $^{68}\text{Ni}$ and to the medium-heavy $^{100,120,132}\text{Sn}$ nuclei. The excitation energies $E^{\ast}$ of the considered compound nuclei 
were calculated and found to increase quickly starting from temperatures $0.5 \leq T \leq 1.0$ MeV. The nucleonic single-particle energies and occupancies change accordingly because of the increasing diffuseness of the Fermi-Dirac distribution with the temperature increase. The dipole response of these nuclei was computed in the FT-RTBA and compared to the finite-temperature relativistic RPA (FT-RRPA).
It was found that the giant dipole resonance (GDR) undergoes additional fragmentation (i) due to the thermal unblocking of the transitions between single-particle states located on the same side of the Fermi surface and (ii) because of the general reinforcement of the particle-vibration coupling (PVC) with the temperature growth.  The low-energy part of the dipole strength distribution is moderately enhanced at temperatures $T \leq  4.0$ MeV and increases dramatically above this temperature range. 
The width of the strength distribution grows rapidly with temperature at $T\geq 1.0$ MeV. The energy-weighted sum rule (EWSR) in a wide finite energy interval remains nearly flat as the temperature increases. The traditional view of the PDR as an oscillation of the weakly bound neutron excess against the isospin-saturated core is nearly maintained up to the temperature $T=5.0$ MeV and changes to the GDR-like pattern above this temperature. The collective behavior of the PDR disappears in the range of temperatures  $1.0\leq T\leq 5.0$ MeV and restores beyond this range which might be, however, already beyond the limit of existence of the nuclei.

\noindent\textbf{Conclusions:} We present a consistent microscopic theory and a numerically stable and executable calculation scheme for computing the nuclear response at finite temperature taking into account the PVC spreading mechanism, in addition to the Landau damping. The presented calculations of the dipole response within a self-consistent relativistic framework reveal that, although the Landau damping plays the leading role in the temperature evolution of the strength distribution, (i)  at moderate temperatures the PVC effects remain almost as strong as at $T=0$ and  (ii) at high temperatures they are tremendously reinforced because of the formation of the new collective low-energy modes. In the dipole channel, the latter effect is responsible for the "disappearance" of the high-frequency GDR or, in other words, brings the GDR to the low-energy domain.
%
\end{abstract}


\maketitle


\section{INTRODUCTION}
The isovector giant dipole resonance (IV GDR) in highly excited nuclei is mainly observed in heavy-ion fusion reactions \cite{Bortignon1998,Harakeh2001}. In the reactions induced by heavy-ion collisions, the fusion between a target of heavy atomic nuclei and a heavy-ion projectile takes place for a long time and a compound nucleus is formed as an intermediate state. During the formation of the compound nucleus, the mean field of the system is established in a very short time and the excitation energy is distributed uniformly among all the single-particle degrees of freedom. Since the time required for the system to achieve the thermal equilibrium is short ($\sim 10^{-22}$ s) compared to the typical time it takes to decay by particle and gamma-ray emission ($\sim 10^{-19}-10^{-9}$ s), one can apply the equilibrium statistical mechanics to describe the hot nucleus in its intermediate states. The nuclear temperature $T$, hence, is assigned using the definition of microcanonical ensemble and related to the excitation energy $E^{\ast}$ as $E^{\ast}\approx aT^{2}$, where $a\approx A/k\;\text{MeV}^{-1}$ is the level density parameter, $k=8-12$ MeV, and $A$ is the mass number \cite{Harakeh2001}. In the Steinwedel-Jensen hydrodynamical model, the IV GDR can be understood as a coherent oscillation of protons against neutrons in the dipole pattern. The general features of the IV GDR built on the excited states can be summarized as follows \cite{Bortignon1998}: 
(i) The EWSR is independent of temperature $T$ and spin angular momentum $J$;
(ii) The centroid energy can be parameterized as $E_{GDR}=18A^{-1/3}+25A^{-1/6}\;\text{MeV}$ and is independent of temperature $T$ and spin angular momentum $J$;
(iii) The width grows with temperature $T$ and spin angular momentum $J$.

The temperature dependence of the high-energy part of the GDR above the neutron emission threshold was extensively studied experimentally in the past \cite{Gaardhoje1984a,GaardhojeEllegaardHerskindEtAl1986,Bracco1989,Ramakrishnan1996,Mattiuzzi1997,Heckman2003}, see also a relatively recent review \cite{Santonocito2006}. In later studies of the dipole response of both ground and excited states of nuclear systems, a concentration of electric dipole strength has been observed in the low-energy region \cite{SavranAumannZilges2013}, being most prominent in neutron-rich nuclei. The distribution of E1 strength below the GDR region is usually classified as pygmy dipole resonance (PDR), which, according to the Steinwedel-Jensen hydrodynamical model, originates from the coherent oscillation of the neutron excess against the isospin-saturated core. Some microscopic models also favor for a collective nature of the PDR which forms at a sufficient amount of the excess neutrons \cite{PaarVretenarKhanEtAl2007,Roca-Maza2018}. There are two important physical aspects related to the study of the PDR. First, the structure of the PDR can significantly enhance the neutron-capture reaction rates of rapid neutron-capture nucleosynthesis (or r-process) \cite{GorielyKhan2002,GorielyKhanSamyn2004,LarsenGoriely2010,LitvinovaLoensLangankeEtAl2009,LitvinovaRingTselyaevEtAl2009}, which is responsible for the formation of chemical elements heavier than iron \cite{Arnould2007}. Second, the PDR can be related to the isovector components of effective nuclear interactions and to the equation of state (EOS) of nuclear matter \cite{PaarVretenarKhanEtAl2007,Roca-Maza2018}. The total PDR strength can provide an experimental constraint on the neutron skin thickness and, in turn, on the symmetry energy of the EOS, which is a key ingredient to study dense astrophysical objects, such as neutron stars \cite{SavranAumannZilges2013}.     

An accurate theoretical description of response of compound nuclei, or nuclei at finite temperature, is an arduous task. In the past, the multipole response of hot nuclei has been studied theoretically within several frameworks, such as  finite-temperature random-phase approximation (FT-RPA) using schematic models \cite{Goodman1981a,Civitarese1984,Vautherin1983,Vautherin1984,Besold1984a,Dang1997} or FT-RPA  with separable forces for deformed rotating nuclei \cite{Faber1983,Gallardo1985}. Approaches beyond FT-RPA include spreading mechanisms and are represented by the finite-temperature nuclear field theory (NFT), which takes into account the coupling between nucleons and low-lying vibrational modes \cite{Bortignon1986,Seva1997}, the collision-integral approach \cite{LacroixChomazAyik1998,LacroixChomazAyik2000}, and the quasiparticle-phonon model (QPM), which operates by the phonon-phonon coupling, formulated as thermofield dynamics \cite{Storozhenko2004}. On the other hand, phenomenological treatment of thermal shape fluctuations and of the particle evaporation have enabled a good description of the overall temperature evolution of the GDR \cite{Alhassid1988,AlhassidBush1990,AlhassidBush1990c,OrmandBortignonBroglia1996,Kusnezov1998}.

The finite-temperature Hartree-Fock-Bogoliubov (FTHFB) equations were derived in \cite{Goodman1981} and  applied for solving the two-level model in \cite{Goodman1981a}. The finite-temperature quasiparticle random phase approximation (FT-QRPA) equations were derived based on FTHFB theory and  solved for a schematic model to calculate the GDR response of hot spherical nuclei \cite{Sommermann1983}. Shortly after that, the formalism was applied successfully to hot rotating nuclei in \cite{RingRobledoEgidoEtAl1984}.  The continuum FT-RPA  \cite{Sagawa1984b} and FT-QRPA \cite{LitvinovaKamerdzhievTselyaev2003,KhanVanGiaiGrasso2004} were successfully applied to various calculations of dipole and quadrupole response of medium-mass nuclei. Later on it was realized that thermal continuum effects may play the major role in explaining the enhancement of the low-energy dipole strength \cite{LitvinovaBelov2013} observed in experiments \cite{VoinovAlginAgvaanluvsanEtAl2004,ToftLarsenBuergerEtAl2011,SimonGuttormsenLarsenEtAl2016}.
More recently, realistic self-consistent approaches in the framework of the relativistic FT-RPA \cite{NiuPaarVretenarEtAl2009} and non-relativistic Skyrme FT-QRPA \cite{YuekselColoKhanEtAl2017} became available for systematic studies of atomic nuclei across the nuclear chart. 

The approaches like RPA and QRPA are commonly classified as the one-loop approximation because they sum only simple ring diagrams. However, correlations beyond this approximation are known to be very important for accurate description of the nuclear response. The above-mentioned  numerical implementations of the finite-temperature approaches beyond R(Q)RPA \cite{Bortignon1986,Seva1997,LacroixChomazAyik2000,Storozhenko2004} are rather limited and mainly focused on the GDR's width problem while the details of the strength distribution, especially of the low-energy strength, are barely addressed. These details, however, become increasingly important now in the context of astrophysical modeling.  Besides that, the results of these approaches are, in some aspects, controversial, although the general framework is, in principle, well established \cite{AdachiSchuck1989,DukelskyRoepkeSchuck1998}. These drawbacks may be related to limited computational capabilities, the use of too simplified nucleon-nucleon interactions and lack of self-consistency. In the present work we aim at building a self-consistent microscopic approach to the finite-temperature nuclear response which (i) is based on the high-quality effective meson-exchange interaction, (ii) takes into account spreading mechanisms microscopically, (iii) is numerically stable and executable, and (iv) allows for systematic studies of both low and high-energy excitations and deexcitations of compound nuclei in a wide range of mass and temperatures. For this purpose, we generalize the response theory developed during the last decade \cite{LitvinovaRingTselyaev2007,LitvinovaRingTselyaev2008,
 LitvinovaRingTselyaev2010,LitvinovaRingTselyaev2013} in the relativistic framework of quantum hadrodynamics for the case of zero temperature.  This approach is based on the covariant energy density functional with the meson-nucleon interaction \cite{Ring1996,VretenarAfanasjevLalazissisEtAl2005} and applies the Green's function formalism and the time blocking approximation \cite{Tselyaev1989} for the time-dependent part of the nucleon-nucleon interaction in the correlated medium. 
 
 The time blocking approximation was formulated originally in Ref. \cite{Tselyaev1989} as a non-perturbative approach to the nuclear response beyond RPA. It is based on the time projection technique within the Green function formalism, which allows for decoupling of configurations of the lowest complexity beyond \textrm{1p1h} (one-particle-one-hole), such as \textrm{1p1h}$\otimes \text{phonon}$ (particle-hole pair coupled to a phonon), from the higher-order ones. This approximation has solved a few conceptual problems at ones: the resulting response function satisfies the general quantum field theory requirements on its analytical properties (locality and unitarity), reduces the Bethe-Salpeter equation to a one-frequency variable equation and ensures a stable numerical scheme for realistic calculations.
 The method was applied systematically in nuclear structure calculations as an extension of the Landau-Migdal theory for non-superfluid nuclear systems \cite{KamerdzhievTertychnyiTselyaev1997} and later generalized for superfluid ones \cite{Tselyaev2007,LitvinovaTselyaev2007}. It has been supplemented by the subtraction procedure introduced in analogy with those of quantum electrodynamics, which enables one to avoid double counting of the particle-vibration coupling (PVC) in the frameworks based on phenomenological mean fields or effective energy density functionals \cite{LitvinovaTselyaev2007,Tselyaev2013}. Since then the time blocking approximation is used consistently in non-relativistic \cite{Lyutorovich2008,Lyutorovich2015,Tselyaev2016,Lyutorovich2018,Lyutorovich2018a} and relativistic \cite{LitvinovaRingTselyaev2007,LitvinovaRingTselyaev2008,
 LitvinovaRingTselyaev2010,LitvinovaRingTselyaev2013,RobinLitvinova2016,RobinLitvinova2018,Robin2019} nuclear structure calculations. The method is systematically improvable and admits extensions which include time-reversed PVC loops as complex ground state correlations \cite{KamerdzhievTertychnyiTselyaev1997,Tselyaev2007,Robin2019} and higher-order configurations \cite{Litvinova2015}. 
 
 At zero temperature the inclusion of the PVC effects in the time blocking approximation leads to a consistent refinement of the calculated spectra in both neutral \cite{LitvinovaRingTselyaevEtAl2009,LitvinovaLoensLangankeEtAl2009,LitvinovaRingTselyaev2010,EndresLitvinovaSavranEtAl2010,
TamiiPoltoratskaNeumannEtAl2011,
MassarczykSchwengnerDoenauEtAl2012,LitvinovaRingTselyaev2013,SavranAumannZilges2013,LanzaVitturiLitvinovaEtAl2014,PoltoratskaFearickKrumbholzEtAl2014,Oezel-TashenovEndersLenskeEtAl2014,EgorovaLitvinova2016} and charge-exchange \cite{MarketinLitvinovaVretenarEtAl2012,LitvinovaBrownFangEtAl2014,RobinLitvinova2016,Litvinova2018,RobinLitvinova2018,Robin2019} channels, as compared to the (Q)RPA approaches, due to the spreading effects. In this work we adopt the Matsubara Green's function formalism for the finite-temperature generalization of the relativistic time blocking approximation (RTBA) \cite{LitvinovaRingTselyaev2007}. The first results obtained within the finite-temperature RTBA (FT-RTBA) were presented in Ref. \cite{LitvinovaWibowo2018} and here we follow up this article with a more detailed formalism and an extended discussion.

The article is organized as follows.  A brief overview of the grand canonical ensemble (GCE) is given in Section II A. In  Section II B, we review the zero-temperature relativistic mean-field (RMF) theory in detail and generalize it for finite temperature.
Section II C is devoted to the general relations defining the finite-temperature response function, while Section II D introduces the finite-temperature time blocking approximation to the particle-vibration coupling amplitude. Extraction of the transition densities is discussed in Section II E.
In Section III, we describe details of the numerical implementation of the developed method and discuss the results of the calculations. The conclusions and outlook are presented in Section IV.

\section{FORMALISM}
\subsection{Grand canonical ensemble}
\label{GCE}
The grand canonical ensemble represents possible states of an open system which can exchange the energy as well as particles with a reservoir and which is characterized by such thermodynamical variables as temperature $T$ and chemical potential $\mu$. For the equilibrium distribution of these states the grand potential \cite{Goodman1981,Sommermann1983,Bellac2004}
\begin{equation}
\Omega=E-TS-\mu N
\label{gp}
\end{equation} 
is minimal, i.e., $\delta\Omega=0$. A positive definite density operator $\hat{\rho}$ is introduced as:
\begin{equation}
\hat{\rho}^{\dag}=\hat{\rho} \;\;\;\;\;\text{and}\;\;\;\;\;\text{Tr}{\hat{\rho}}=1,
\end{equation}   
where the symbol $\text{Tr}$ represents a summation of all diagonal elements of the matrix or matrices under the operation, and the summation includes all possible numbers of particles of all kinds and all possible states of these particles. In terms of the density operator $\hat{\rho}$, the average energy $E$, the average particle number $N$, and the entropy $S$ of the system can be determined as
\begin{eqnarray}
E&=&\langle\hat{H}\rangle=\text{Tr}(\hat{\rho}\hat{H}),\\
N&=&\langle\hat{N}\rangle=\text{Tr}(\hat{\rho}\hat{N}),\\
S&=&\langle -k\ln\hat{\rho}\rangle=-k\text{Tr}(\hat{\rho}\ln\hat{\rho}),
\end{eqnarray}
where $k$ is the Boltzmann's constant. From the last three equations and the constraint $\text{Tr}\hat{\rho}=1$, the minimization of grand potential $\Omega$ leads to 
\begin{equation}
\text{Tr}\left\{\delta\hat{\rho}\left[\hat{H}-\mu\hat{N}+kT(\ln\hat{\rho}+1)\right]\right\}=0.
\end{equation}
Since $\delta\hat{\rho}$ is arbitrary, the last equation and the constraint $\text{Tr}\hat{\rho}=1$ gives the solution for the density operator $\hat{\rho}$ of the form
\begin{eqnarray}
\label{Density Operator}\hat{\rho}&=&Z^{-1}e^{-(\hat{H}-\mu\hat{N})/kT},\\
\label{Grand Partition Function}Z&=&\text{Tr}\left[e^{-(\hat{H}-\mu\hat{N})/kT}\right],
\end{eqnarray}
where $Z$ is the grand partition function. The thermal average of an operator $\hat{\mathcal{O}}$ then reads: 
\begin{equation}
\langle\hat{\mathcal{O}}\rangle=\text{Tr}(\hat{\rho}\hat{\mathcal{O}})=Z^{-1}\text{Tr}\left[e^{-(\hat{H}-\mu\hat{N})/kT}\hat{\mathcal{O}}\right].
\end{equation} 
Given the grand canonical partition function $Z$, several thermodynamic quantities can be obtained as follows \cite{Bellac2004}:
\begin{eqnarray}
\Omega&=&-kT\ln Z=-PV,\\
E&=&-\left.\frac{\partial\ln Z}{\partial\beta}\right|_{z,V},\\
N&=&z\left.\frac{\partial\ln Z}{\partial z}\right|_{\beta,V},\\
S&=&\frac{1}{T}\left[E-\mu N-\Omega\right],
\end{eqnarray} 
where $z=e^{\mu/kT}$ is the fugacity and $\beta=1/kT$. In the following we take the value of $k=1$.

\subsection{Relativistic mean-field theory at zero and finite temperatures}
\label{FT-RMF}
We start with the Lagrangian density of quantum hadrodynamics (QHD) \cite{Serot1986,Ring1996, Meng2016}: 
\begin{eqnarray}
\label{QHD Lagrangian}\mathcal{L}&=&\overline{\psi}(i\gamma^{\mu}\partial_{\mu}-M)\psi+\frac{1}{2}\left(\partial^{\mu}\sigma\partial_{\mu}\sigma-m^{2}_{\sigma}\sigma^{2}\right)-\nonumber\\
&-&\frac{1}{2}\left(\frac{1}{2}\Omega_{\mu\nu}\Omega^{\mu\nu}-m_{\omega}^{2}\omega_{\mu}\omega^{\mu}\right)-\nonumber\\
&-&\frac{1}{2}\left(\frac{1}{2}\vec{R}_{\mu\nu}\vec{R}^{\mu\nu}-m_{\rho}^{2}\vec{\rho}_{\mu}\vec{\rho}^{\mu}\right)-\frac{1}{4}F^{\mu\nu}F_{\mu\nu}-\nonumber\\
&-&\overline{\psi}\Gamma_{\sigma}\sigma\psi-\overline{\psi}\Gamma_{\omega}^{\mu}\omega_{\mu}\psi-\overline{\psi}\vec{\Gamma}^{\mu}_{\rho}\vec{\rho}_{\mu}\psi\nonumber\\
&-&\overline{\psi}\Gamma_{e}^{\mu}A_{\mu}\psi-U(\sigma),
\end{eqnarray}
where $M$ is the mass of the nucleon, $\psi$ is the nucleonic field, $\sigma$ is the scalar $\sigma$-meson, $m_{a}$ ($a=\sigma,\;\omega,\;\rho$) are meson masses, and the tensors $\Omega^{\mu\nu}$, $\vec{R}^{\mu\nu}$, and $F^{\mu\nu}$ represent the $\omega$-meson, $\rho$-meson and the electromagnetic field, respectively, 
\begin{eqnarray}
\Omega^{\mu\nu}&=&\partial^{\mu}\omega^{\nu}-\partial^{\nu}\omega^{\mu},\\
\vec{R}^{\mu\nu}&=&\partial^{\mu}\vec{\rho}^{\nu}-\partial^{\nu}\vec{\rho}^{\mu},\\
F^{\mu\nu}&=&\partial^{\mu}A^{\nu}-\partial^{\nu}A^{\mu}.
\end{eqnarray}
We use the arrow to denote isovectors and boldface letters to indicate vectors in three-dimensional space. The Greek indices run over the components in Minkowski space: 0, 1, 2, and 3, where 0 represents the time-like component and the other denote the space-like components. We also apply the Einstein summation convention, i.e., summation over the repeated indices is implied. The meson-nucleon vertices $\Gamma_{\sigma}$, $\Gamma_{\omega}^{\mu}$, $\vec{\Gamma}_{\rho}^{\mu}$, and photon-nucleon vertex $\Gamma_{e}^{\mu}$ read
\begin{equation}
\label{Vertices}\Gamma_{\sigma}=g_{\sigma},\;\;\Gamma^{\mu}_{\omega}=g_{\omega}\gamma^{\mu},\;\;\vec{\Gamma}_{\rho}^{\mu}=g_{\rho}\gamma^{\mu}\vec{\tau},\;\;\Gamma^{\mu}_{e}=\frac{1}{2}(1+\tau_{3})e\gamma^{\mu},
\end{equation}
where $g_{a}$ ($a=\sigma,\;\omega,\;\rho$) and $e$ are the corresponding coupling constants. The non-linear $\sigma$ self-interaction term $U(\sigma)$, following Ref. \cite{BogutaBodmer1977}, reads:
\begin{equation}
U(\sigma)=\frac{1}{3}g_{2}\sigma^{3}+\frac{1}{4}g_{3}\sigma^{4}.
\end{equation}
The corresponding Euler-Lagrange equations are, 
for the nucleonic fields:
\begin{equation}
\label{Nucleon Field Equation}\left[i\gamma_{\mu}\partial^{\mu}-M-\sum_{m}\Gamma_{m}\phi_{m}(\textbf{r},t)\right]\psi(\textbf{r},t)=0,
\end{equation}
where $m=\{\sigma,\;\omega,\;\rho,\;e\}$, $\Gamma_{m}=\left\{\Gamma_{\sigma},\;\Gamma^{\mu}_{\omega},\;\vec{\Gamma}^{\mu}_{\rho},\;\Gamma^{\mu}_{e}\right\}$, $\phi_{m}=\left\{\sigma,\;\omega^{\mu},\;\vec{\rho}^{\mu},\;A^{\mu}\right\}$, and for the meson and electromagnetic fields:
\begin{eqnarray}
\label{Sigma-Meson Field}(\square+m_{\sigma}^{2})\sigma(\textbf{r},t)&=&-\overline{\psi}(\textbf{r},t)\Gamma_{\sigma}\psi(\textbf{r},t)-\frac{dU(\sigma)}{d\sigma},\\
\label{Omega-Meson Field}(\square+m_{\omega}^{2})\omega^{\mu}(\textbf{r},t)&=&\overline\psi(\textbf{r},t)\Gamma_{\omega}^{\mu}\psi(\textbf{r},t),\\
\label{Rho-Meson Field}(\square+m_{\rho}^{2})\vec{\rho}^{\mu}(\textbf{r},t)&=&\overline\psi(\textbf{r},t){\vec\Gamma}_{\rho}^{\mu}\psi(\textbf{r},t),\\
\label{Photon Field}\square A^{\mu}(\textbf{r},t)&=&\overline{\psi}(\textbf{r},t)\Gamma_{e}^{\mu}\psi(\textbf{r},t),
\end{eqnarray}
where 
\begin{equation}
\square:=\partial_{\mu}\partial^{\mu}=\frac{\partial^{2}}{\partial t^{2}}-\nabla^{2}
\end{equation}
is the d'Alembertian operator and we have imposed the Lorenz gauge condition:  
\begin{equation}
\partial_{\mu}\omega^{\mu}(\textbf{r},t)=0,\;\;\;\partial_{\mu}\vec{\rho}^{\mu}(\textbf{r},t)=0,\;\;\;\partial_{\mu}A^{\mu}(\textbf{r},t)=0.
\end{equation}

\noindent To solve the time-dependent self-consistent field equations, i.e., Eqs. \eqref{Nucleon Field Equation}-\eqref{Photon Field}, directly is a very non-trivial task. The leading  approximation implies that the meson field and the electromagnetic field operators are replaced by their expectation values in the nuclear ground state, which constitutes the relativistic mean-field (RMF) approximation. As a result, the nucleons move independently in the classical meson fields. In this work, we will use the RMF as a basis for the nuclear response calculations by means of the solution of the Bethe-Salpeter equation, where we approximately restore the time dependence neglected in the RMF.

The corresponding covariant energy density functional (CEDF) is given as
\begin{eqnarray}
E_{\text{RMF}}[\hat{\rho},\phi]&=&\langle\Phi|\hat{H}_{\text{RMF}}|\Phi\rangle\nonumber\\
&=&\text{Tr}\left[(\boldsymbol{\alpha}\cdot\textbf{p}+\beta M+\beta\Gamma_{m}\phi_{m})\hat{\rho}\right]+\nonumber\\
&+&\frac{1}{2}\int d^{3}\textbf{r}\left[\dot{\sigma}^{2}+(\nabla\sigma)^{2}+m^{2}_{\sigma}\sigma^{2}\right]-\nonumber\\
&-&\frac{1}{2}\int d^{3}\textbf{r}\left[\dot{\omega}^{j}_{}\dot{\omega}_{j}+(\boldsymbol{\nabla}\omega^{\mu})\cdot(\boldsymbol{\nabla}\omega_{\mu})+m^{2}_{\omega}\omega^{\mu}\omega_{\mu}\right]-\nonumber\\
&-&\frac{1}{2}\int d^{3}\textbf{r}\left[\dot{\vec{\rho}}^{j}\dot{\vec{\rho}}_{j}+(\boldsymbol{\nabla}\vec{\rho}^{\mu})\cdot(\boldsymbol{\nabla}\vec{\rho}_{\mu})+m^{2}_{\rho}\vec{\rho}^{\mu}\vec{\rho}_{\mu}\right]-\nonumber\\
&-&\frac{1}{2}\int d^{3}\textbf{r}\left[\dot{A}^{j}\dot{A}_{j}+(\boldsymbol{\nabla}A^{\mu})\cdot(\boldsymbol{\nabla}A_{\mu})\right]+\nonumber\\
&+&\int d^{3}\textbf{r}U(\sigma),
\label{cedf}
\end{eqnarray} 
where $\beta=\gamma^{0}$, $\boldsymbol{\alpha}=\beta\boldsymbol{\gamma}$, and the latin index $j$ denotes only the space-like components. To obtain Eq. (\ref{cedf}), we have considered the zero-components of the vector fields, i.e., $\omega^{0}(\textbf{r}),\;\vec{\rho}^{0}(\textbf{r}),\;A^{0}(\textbf{r})$, being static, or time-independent. 
This consideration is valid for the case of heavy mesons and long-wave photons, for which the one-meson and one-photon exchange potentials are nearly time-independent and take the forms of static Yukawa and Coulomb potentials, respectively 
\cite{Baranger1975, Serot1986}.  
The trace in Eq. (\ref{cedf}) represents a sum over Dirac indices of the density matrix $\hat{\rho}$ and an integral in the coordinate space, and we have used the  mean-field approximation for the meson and electromagnetic fields:
\begin{equation}
\langle\Phi|\phi_{m}|\Phi\rangle =\phi_{m},\ \ \ \ 
\langle\Phi|\phi_{m}\hat{a}^{\dag}_{k'}\hat{a}_{k}|\Phi\rangle =\phi_{m}\langle\Phi|\hat{a}^{\dag}_{k'}\hat{a}_{k}|\Phi\rangle.
\end{equation}

The finite-temperature generalization of the RMF theory can be established using the general prescription given in subsection \ref{GCE}. 
Minimization of the grand potential (\ref{gp}) with the CEDF of Eq. ({\ref{cedf}}) leads to the single-particle density operator of the form:
\begin{eqnarray}
\label{RMF Formula of Density Operator}\hat{\rho}&=&\frac{e^{-(\hat{\mathcal{H}}-\mu\hat{\mathcal{N}})/T}}{\text{Tr}\left[e^{-(\hat{\mathcal{H}}-\mu\hat{\mathcal{N}})/T}\right]},
\end{eqnarray}
where the Hamilton operator $\hat{\mathcal{H}}$ is given by
\begin{equation}
\hat{\mathcal{H}}=\boldsymbol{\alpha}\cdot\textbf{p}+\beta\left[M+\Gamma_{m}\phi_{m}(\textbf{r},t)\right].
\end{equation}
The expression of operator $\hat{\mathcal{H}}$ can be simplified further for the stationary solutions: 
\begin{equation}
\label{Single Particle Wave Function}\psi_{k}(\textbf{r},t)=\psi_{k}(\textbf{r})e^{-i\varepsilon_{k} t},
\end{equation} 
where $\varepsilon_{k}$ is the single-particle energy of the state $k$. The Dirac equation (\ref{Nucleon Field Equation}) now becomes
\begin{equation}
\label{Basis for Dirac Hamiltonian}\hat{h}^{D}\psi_{k}(\textbf{r})=\varepsilon_{k}\psi_{k}(\textbf{r}),
\end{equation}
where $\hat{h}^{D}$ is the Dirac Hamiltonian:
\begin{equation}
\hat{h}^{D}=\boldsymbol{\alpha}\cdot\textbf{p}+\beta(M+\widetilde{\Sigma}(\textbf{r})),
\end{equation}
and $\widetilde{\Sigma}(\textbf{r})$ is the static RMF self-energy (mass operator):
\begin{equation}
\widetilde{\Sigma}(\textbf{r})=\sum_{m}\Gamma_{m}\phi_{m}(\textbf{r}).
\end{equation}
Further we assume the time-reversal symmetry of the RMF, so that the current densities are equal to zero and, therefore, the spatial components of the meson and electromagnetic fields vanish. Furthermore, the isospin $\tau_{3}$ is supposed to be a good quantum number, so that the only non-zero component of $\vec{\rho}^{0}$ is $\rho_{3}^{0}$. Thus, the static RMF self-energy is decomposed into the scalar $\widetilde{\Sigma}_{s}(\textbf{r})$ and the vector time-like  $\widetilde{\Sigma}^{0}(\textbf{r})$ components as follows:   
\begin{eqnarray}
\widetilde{\Sigma}_{s}(\textbf{r})&=&g_{\sigma}\sigma(\textbf{r}),\\
\widetilde{\Sigma}^{0}(\textbf{r})&=&\beta\left[g_{\omega}\omega^{0}(\textbf{r})+\frac{1}{2}(1+\tau_{3})eA^{0}(\textbf{r})+g_{\rho}\tau_{3}\rho_{3}^{0}\right].
\end{eqnarray}
The non-vanishing meson and electromagnetic fields satisfy the following equations:
\begin{eqnarray}
\label{RMF Equation for Sigma Meson}\left(-\nabla^{2}+m^{2}_{\sigma}\right)\sigma(\textbf{r})&=&-g_{\sigma}\rho_{s}(\textbf{r})-\frac{dU(\sigma)}{d\sigma},\\
\label{RMF Equation for Omega Meson}\left(-\nabla^{2}+m^{2}_{\omega}\right)\omega^{0}(\textbf{r})&=&g_{\omega}\rho_{v}(\textbf{r}),\\
\label{RMF Equation for Rho Meson}\left(-\nabla^{2}+m^{2}_{\rho}\right)\rho^{0}_{3}(\textbf{r})&=&g_{\rho}\rho_{3}(\textbf{r}),\\
\label{RMF Equation for Photon}-\nabla^{2}A^{0}(\textbf{r})&=&e\rho_{c}(\textbf{r}),
\end{eqnarray}
where $\rho_{s},\;\rho_{v},\;\rho_{3},$ and $\rho_{c}$ respectively are the scalar, baryon, isovector, and charge densities:
\begin{eqnarray}
\label{Scalar Density}\rho_{s}(\textbf{r},t)&=&\sum_{k\ell}\rho_{\ell k}\overline{\psi}_{k}(\textbf{r})\psi_{\ell}(\textbf{r})e^{i(\varepsilon_{k}-\varepsilon_{\ell})t},\\
\label{Baryon Density}\rho_{v}(\textbf{r},t)&=&\sum_{k\ell}\rho_{\ell k}\psi^{\dag}_{k}(\textbf{r})\psi_{\ell}(\textbf{r})e^{i(\varepsilon_{k}-\varepsilon_{\ell})t},\\
\label{Isovector Density}\rho_{3}(\textbf{r},t)&=&\sum_{k\ell}\rho_{\ell k}\psi^{\dag}_{k}(\textbf{r})\tau_{3}\psi_{\ell}(\textbf{r})e^{i(\varepsilon_{k}-\varepsilon_{\ell})t},\\
\label{Charge}\rho_{c}(\textbf{r},t)&=&\sum_{k\ell}\rho_{\ell k}\psi^{\dag}_{k}(\textbf{r})\frac{1}{2}\left(1+\tau_{3}\right)\psi_{\ell}(\textbf{r})e^{i(\varepsilon_{k}-\varepsilon_{\ell})t}.
\end{eqnarray}
At zero temperature, the single-particle density matrix $\rho_{\ell k}=\delta_{\ell k}$ for states $|\textrm{h}\rangle$ below the Fermi level and zero otherwise. Therefore, the densities (\ref{Scalar Density})-(\ref{Charge}) reduce to:
\begin{eqnarray}
\rho_{s}(\textbf{r})&=&\sum_{k=1}^A\overline{\psi}_{k}(\textbf{r})\psi_{k}(\textbf{r}),\\
\rho_{v}(\textbf{r})&=&\sum_{k=1}^A\psi^{\dag}_{k}(\textbf{r})\psi_{k}(\textbf{r}),\\
\rho_{3}(\textbf{r})&=&\sum_{k=1}^A\psi^{\dag}_{k}(\textbf{r})\tau_{3}\psi_{k}(\textbf{r}),\\
\rho_{c}(\textbf{r})&=&\sum_{k=1}^A\psi^{\dag}_{k}(\textbf{r})\frac{1}{2}(1+\tau_{3})\psi_{k}(\textbf{r}).
\end{eqnarray}

Under the above-mentioned assumptions the operator $\hat{\mathcal{H}}$ becomes the Dirac Hamiltonian $\hat{h}^{D}$ which, in the basis of Eq. (\ref{Basis for Dirac Hamiltonian}), can be written as 
\begin{equation}
\hat{h}^{D}=\sum_{k}\varepsilon_{k}\hat{a}^{\dag}_{k}\hat{a}_{k},
\end{equation}
while the total particle number operator $\hat{\mathcal{N}}$ can be expressed as follows:
\begin{equation}
\hat{\mathcal{N}}=\sum_{k}\hat{a}^{\dag}_{k}\hat{a}_{k}.
\label{N1}
\end{equation}
From the last two equations, we obtain the grand partition function $\mathcal{Z}$:
\begin{equation}
\mathcal{Z}=\prod_{k}\left[1+ze^{-\varepsilon_{k}/T}\right]
\end{equation}
and the mean value of the operator $\hat{\mathcal{N}}$:
\begin{equation}
\mathcal{N}=\langle\hat{\mathcal{N}}\rangle=\sum_{k}n_{k},
\label{N2}
\end{equation}
while the Fermi-Dirac occupation number $n_{k}$ of the state $k$ reads:
\begin{equation}
n_k(T) = n(\varepsilon_{k},T)=\frac{1}{1+e^{(\varepsilon_{k}-\mu)/T}},
\label{FermiDirac} 
\end{equation}
where  $\sum_{k}n_{k}=A$ with $A$ being the total number of nucleons. From Eqs. (\ref{N1},\ref{N2}) we can obtain that
\begin{equation}
\langle\hat{a}^{\dag}_{k}\hat{a}_{k}\rangle=n_{k}.
\end{equation}
At finite temperature, the single-particle density matrix $\rho_{k\ell}$ becomes \cite{Goodman1981,Abrikosov1965}
\begin{equation}
\label{FT Rho kl}\rho_{k\ell}=\text{Tr}\left[\hat{\rho}\hat{a}_{\ell}^{\dag}\hat{a}_{k}\right]=\langle\hat{a}_{\ell}^{\dag}\hat{a}_{k}\rangle=\delta_{\ell k}n_{k}
\end{equation} 
and the densities (\ref{Scalar Density})-(\ref{Charge}) reduce to the following set:
\begin{eqnarray}
\label{FT-Scalar}\rho_{s}(\textbf{r})&=&\sum_{k}n_{k}\overline{\psi}_{k}(\textbf{r})\psi_{k}(\textbf{r}),\\
\label{FT-Baryon}\rho_{v}(\textbf{r})&=&\sum_{k}n_{k}\psi^{\dag}_{k}(\textbf{r})\psi_{k}(\textbf{r}),\\
\label{FT-Isovector}\rho_{3}(\textbf{r})&=&\sum_{k}n_{k}\psi^{\dag}_{k}(\textbf{r})\tau_{3}\psi_{k}(\textbf{r}),\\
\label{FT-Charge}\rho_{c}(\textbf{r})&=&\sum_{k}n_{k}\psi^{\dag}_{k}(\textbf{r})\frac{1}{2}(1+\tau_{3})\psi_{k}(\textbf{r}).
\end{eqnarray}


In the present work we will deal with spherical nuclei, which implies spherical symmetry, so that the spinor $\psi_{k}(\textbf{r})$ is specified by the set of quantum numbers $k=\{(k),m_{k}\}$, where $(k)=\{n_{k},\;j_{k},\;\pi_{k},\;\tau_{k}\}$. Here $n_{k}$ is the radial quantum number, $j_{k},\;m_{k}$ are the angular momentum quantum number and its z-component, respectively, $\pi_{k}$ is the parity, and $\tau_{k}$ is the isospin. Taking into account the spin $s$ and isospin $t$ variables, the Dirac spinor $\psi_{k}(\textbf{r},s,t)$ takes the form \cite{LitvinovaRingTselyaev2007}:
\begin{eqnarray}
\psi_{k}(\textbf{r},s,t)&=&\left( \begin{array}{c}
f_{(k)}(r)\Phi_{l_{k}j_{k}m_{k}}(\vartheta,\varphi,s) \\ 
ig_{(k)}(r)\Phi_{\tilde{l}_{k}j_{k}m_{k}}(\vartheta,\varphi,s) \\ 
\end{array}  \right)\chi_{\tau_{k}}(t),
\end{eqnarray}
where the orbital angular momenta of the large and small components, i.e., $l_{k}$ and $\tilde{l}_{k}$, respectively, are related to the parity $\pi_{k}$ as follows:
\begin{equation}
\left\{ \begin{array}{ccc}
l_{k}=j_{k}+\tfrac{1}{2},\;\tilde{l}_{k}=j_{k}-\tfrac{1}{2} & \text{for} & \pi_{k}=(-1)^{j_{k}+\tfrac{1}{2}} \\ 
l_{k}=j_{k}-\tfrac{1}{2},\;\tilde{l}_{k}=j_{k}+\tfrac{1}{2} & \text{for} & \pi_{k}=(-1)^{j_{k}-\tfrac{1}{2}}
\end{array},  \right.
\end{equation}
$f_{(k)}(r)$ and $g_{(k)}(r)$ are radial wave functions, and $\Phi_{ljm}(\vartheta,\varphi,s)$ is the spin-angular part:
\begin{equation}
\Phi_{ljm}(\vartheta,\varphi,s)=\sum_{m_{s}m_{l}}\langle\tfrac{1}{2}m_{s}lm_{l}|jm\rangle Y_{lm_{l}}(\vartheta,\varphi)\chi_{m_{s}}(s).
\end{equation}

\subsection{Finite-temperature response function}
\label{ft-response}
The main observable under consideration will be the strength function which describes the probability distribution of nuclear transitions under a weak perturbation induced by an external field $\hat{V}^{0}$. At zero temperature, it is defined as \cite{KamerdzhievTertychnyiTselyaev1997}:
\begin{eqnarray}
S(\omega)=\sum_{n>0}\Big[|\langle n|\hat{V}^{0\dag}|0\rangle|^{2}&&\delta(\omega-\omega_{n})-\nonumber\\
&&-|\langle n|\hat{V}^{0}|0\rangle|^{2}\delta(\omega+\omega_{n})\Big],
\label{sf}
\end{eqnarray}
where $\omega_{n}=E_{n}-E_{0}$ is the excitation energy with respect to the ground state energy $E_{0}$. The states $|n\rangle$ and the energies $E_{n}$ are the exact eigenstates and eigenvalues of the many-body Hamiltonian $\hat{H}$ characterized by a set of quantum numbers $n$. As possible external fields $\hat{V}^{0}$ we will consider operators of the one-body character:
\begin{equation}
\hat{V}^{0}=\sum_{k_{1}k_{2}}V^{0}_{k_{1}k_{2}}\hat{a}^{\dag}_{k_{1}}\hat{a}_{k_{2}}.
\end{equation} 
The transition density between the ground state and the excited state $n$ is given by: 
\begin{equation}
\rho^{n0}_{k_{1}k_{2}}=\langle n|\hat{a}_{k_{1}}^{\dag}\hat{a}_{k_{2}}|0\rangle ,
\end{equation}
which differs by the complex conjugation from that of Ref. \cite{LitvinovaRingTselyaev2007}.
At zero temperature, the response function $R(\omega)$, defined as \cite{KamerdzhievTertychnyiTselyaev1997}
\begin{eqnarray}
R_{k_{1}k_{2},k_{3}k_{4}}(\omega)=\int_{-\infty}^{\infty}\frac{d\varepsilon}{2\pi i}R_{k_{1}k_{2},k_{3}k_{4}}(\omega,\varepsilon)\nonumber\\
= \sum_{n>0}\left[\frac{\rho_{k_{2}k_{1}}^{n0\ast}\rho_{k_{4}k_{3}}^{n0}}{\omega+\omega_{n}+i\delta}-\frac{\rho_{k_{1}k_{2}}^{n0}\rho^{n0\ast}_{k_{3}k_{4}}}{\omega-\omega_{n}+i\delta}\right], \ \ \ \ \ \ \ \delta\rightarrow+0,
\end{eqnarray}
completely determines the strength function $S(\omega)$ via:
\begin{equation}
S(\omega)=\frac{1}{\pi}\lim_{\Delta\rightarrow+0}\text{Im}\Pi_{pp}(\omega+i\Delta),
\label{strength}
\end{equation}
where the polarizability $\Pi_{pp}(\omega)$ is defined as the double convolution of the full response function $R(\omega)$ with the external field $\hat{V}^{0}$:
\begin{equation}
\Pi_{pp}(\omega)=\sum_{k_{1}k_{2}k_{3}k_{4}}V^{0\ast}_{k_{2}k_{1}}R_{k_{1}k_{2},k_{3}k_{4}}(\omega)V^{0}_{k_{4}k_{3}}.
\label{polarizability}
\end{equation}
The quantity $\Delta$ in Eq. (\ref{strength}) is a finite imaginary part of the energy variable, which is commonly used as a smearing parameter related to the finite experimental resolution and missing microscopic effects. 

The strength function at finite temperature is defined as \cite{RingRobledoEgidoEtAl1984}
\begin{eqnarray}
\tilde{S}(E)&=&\sum_{if}p_{i}|\langle f|\hat{V}^{0\dag}|i\rangle|^{2}\delta(E-E_{f}+E_{i})\equiv S_{+}(E),
\end{eqnarray}
where $f$ denotes the set of final states and $i$ stands for possible initial states distributed with the probabilities $p_{i}$:
\begin{equation}
p_{i}=\frac{e^{-E_{i}/T}}{\sum_{j}e^{-E_{j}/T}}.
\end{equation}
Using the principle of detailed balance, for the emission strength function we have: 
\begin{equation}
S_{-}(E)=\sum_{if}p_{i}|\langle f|\hat{V}^{0}|i\rangle|^{2}\delta(E+E_{f}-E_{i})=e^{-E/T}S_{+}(E),
\end{equation}
so that the strength function $\tilde{S}(E)$ becomes
\begin{eqnarray}
\tilde{S}(E)&=&\frac{1}{1-e^{-E/T}}\left[S_{+}(E)-S_{-}(E)\right]\nonumber\\
&=&\lim_{\Delta\rightarrow+0}\frac{1}{\pi}\frac{1}{1-e^{-E/T}}\text{Im}\left[\sum_{if}p_{i}\left\{\frac{|\langle f|\hat{V}^{0}|i\rangle|^{2}}{E+E_{f}-E_{i}+i\Delta}-\right.\right.\nonumber\\
&-&\left.\left.\frac{|\langle f|\hat{V}^{0\dag}|i\rangle|^{2}}{E-E_{f}+E_{i}+i\Delta}\right\}\right].
\end{eqnarray}
In terms of the finite-temperature response function ($\delta\rightarrow+0$):
\begin{eqnarray}
\mathcal{R}_{k_{1}k_{2},k_{3}k_{4}}(E)&=&\sum_{if}p_{i}\left\{\frac{\langle f|\hat{a}^{\dag}_{k_{4}}\hat{a}_{k_{3}}|i\rangle\langle i|\hat{a}^{\dag}_{k_{1}}\hat{a}_{k_{2}}|f\rangle}{E+E_{f}-E_{i}+i\delta}-\right.\nonumber\\
&-&\left.\frac{\langle f|\hat{a}^{\dag}_{k_{1}}\hat{a}_{k_{2}}|i\rangle\langle i|\hat{a}^{\dag}_{k_{4}}\hat{a}_{k_{3}}|f\rangle}{E-E_{f}+E_{i}+i\delta}\right\},
\end{eqnarray}
the strength function $\tilde{S}(E)$ can be expressed as
\begin{eqnarray}
\label{Strength}
\tilde{S}(E)&=&\frac{1}{1-e^{-E/T}}{S}(E),\\
{S}(E)&=&\lim_{\Delta\rightarrow+0}\frac{1}{\pi}\text{Im}\sum_{k_{1}k_{2}k_{3}k_{4}}V^{0\ast}_{k_{2}k_{1}}\mathcal{R}_{k_{1}k_{2},k_{3}k_{4}}(E+i\Delta)V^{0}_{k_{4}k_{3}}.\nonumber\\
\end{eqnarray}
The factor $\left[1-\exp(-E/T)\right]^{-1}$  is, thereby, the new feature which is inherent for the finite-temperature strength function. In particular, it influences the low-energy behavior of $\tilde{S}(E)$, in addition to the appearance of new poles in the response function, and makes the zero-energy limit of $\tilde{S}(E)$ finite at $T>0$, in contrast to that of the spectral density  ${S}(E)$, whose zero-energy limit is zero at all temperatures.

In analogy to the case of zero temperature \cite{KamerdzhievTertychnyiTselyaev1997}, the spectral representation of the finite-temperature response function can be defined as
\begin{equation}
\mathcal{R}_{k_{1}k_{2},k_{3}k_{4}}(\omega_{n}):=T\sum_{\ell}\mathcal{R}_{k_{1}k_{2},k_{3}k_{4}}(\omega_{n},\varepsilon_{\ell}),
\end{equation}
via the solution of the Bethe-Salpeter equation (BSE)
in the particle-hole (\textrm{ph}) channel:
\begin{eqnarray}
\label{BSE}&\mathcal{R}&_{k_{1}k_{2},k_{3}k_{4}}(\omega_{n},\varepsilon_{\ell})\nonumber\\
&=&\mathcal{R}^{0}_{k_{1}k_{2},k_{3}k_{4}}(\omega_{n},\varepsilon_{\ell})-\sum_{k_{5}k_{6}k_{7}k_{8}}\mathcal{R}^{0}_{k_{1}k_{2},k_{5}k_{6}}(\omega_{n},\varepsilon_{\ell})\times\nonumber\\
&\times&T\sum_{\ell'}\mathcal{U}_{k_{5}k_{6},k_{7}k_{8}}(\omega_{n},\varepsilon_{\ell},\varepsilon_{\ell'})\mathcal{R}_{k_{7}k_{8},k_{3}k_{4}}(\omega_{n},\varepsilon_{\ell'}),
\end{eqnarray}
where $\mathcal{R}^{0}_{k_{1}k_{2},k_{3}k_{4}}(\omega_{n},\varepsilon_{\ell})=-\mathcal{G}_{k_{3}k_{1}}(\omega_{n}+\varepsilon_{\ell})\mathcal{G}_{k_{2}k_{4}}(\varepsilon_{\ell})$ is the uncorrelated particle-hole response, $\mathcal{G}$ is the exact one-body Matsubara Green's function, and $\mathcal{U}$ is the nucleon-nucleon interaction amplitude. In this work we will consider nuclear response in the particle-hole channel, so that, accordingly, $\mathcal{U}$ will be the particle-hole (direct-channel) interaction. In the following, where there is no need of specifying the basis indices and energy arguments, we will use relations in the operator form. In particular, the BSE (\ref{BSE}) reads: 
\begin{equation}
\mathcal{R}=\mathcal{R}^{0}-\mathcal{R}^{0}\mathcal{U}\mathcal{R},
\end{equation}
with $\mathcal{R}^{0}=-\mathcal{G}\mathcal{G}$.
The Matsubara frequencies $\omega_{n}$, $\varepsilon_{\ell}$, and $\varepsilon_{\ell'}$ are discrete variables defined as $\omega_{n}=2n\pi T$, $\varepsilon_{\ell}=(2\ell+1)\pi T$, and $\varepsilon_{\ell'}=(2\ell'+1)\pi T$, where $n$, $\ell$, and $\ell'$ are integer numbers. The Green's function $\mathcal{G}$ satisfies the Dyson equation
\begin{equation}
\label{Exact Dyson}\mathcal{G}_{k_{1}k_{2}}(\varepsilon_{\ell})=\mathcal{G}^{0}_{k_{1}k_{2}}(\varepsilon_{\ell})+\sum_{k_{3}k_{4}}\mathcal{G}^{0}_{k_{1}k_{3}}(\varepsilon_{\ell})\Sigma_{k_{3}k_{4}}(\varepsilon_{\ell})\mathcal{G}_{k_{4}k_{2}}(\varepsilon_{\ell}),
\end{equation}
where $\mathcal{G}^{0}$ is the unperturbed one-body Matsubara Green's function and $\Sigma$ is the self-energy (mass operator). In the general equation of motion (EOM) approach \cite{VinhMau1979,AdachiSchuck1989,DukelskyRoepkeSchuck1998}, the self-energy $\Sigma$ decomposes into the static and the time-dependent parts, and the latter translates to the energy-dependent term in the energy domain:
\begin{equation}
\Sigma(\varepsilon)=\widetilde{\Sigma}+\Sigma^{e}(\varepsilon),
\end{equation} 
where $\Sigma^{e}$ represents the energy-dependent, in general, non-local self-energy. 
It is convenient to introduce the temperature mean-field Green's function $\widetilde{\mathcal{G}}$, such as 
\begin{equation}
\label{Mean-Field Dyson}\widetilde{\mathcal{G}}=\mathcal{G}^{0}+\mathcal{G}^{0}\widetilde{\Sigma}\widetilde{\mathcal{G}}, \ \ \ \ \ \ 
\mathcal{G}=\widetilde{\mathcal{G}}+\widetilde{\mathcal{G}}\Sigma^{e}\mathcal{G}.
\end{equation}
In the imaginary-time $\tau$ ($0<\tau<1/T$) representation, the temperature mean-field Green's function $\widetilde{\mathcal{G}}$ reads \cite{Abrikosov1965}:
\begin{eqnarray}
\label{Full Mean-Field}\widetilde{\mathcal{G}}(2,1)&=&\sum_{\sigma}\widetilde{\mathcal{G}}^{\sigma}(2,1),\\
\label{Component Mean-Field}\widetilde{\mathcal{G}}^{\sigma}(2,1)&=&-\sigma\delta_{k_{1}k_{2}}n[-\sigma\varepsilon_{k_{1}},T]\times\nonumber\\
&\times&e^{-(\varepsilon_{k_{1}}-\mu)\tau_{21}}\theta(\sigma\tau_{21}),
\end{eqnarray}
where $\tau_{21}=\tau_{2}-\tau_{1}$ ($-1/T<\tau_{21}<1/T$), $\theta(\tau)$ is the Heaviside step-function, and each number index represents all single-particle quantum numbers and imaginary-time variable: $1=\{k_{1},\tau_{1}\}$. The index $\sigma=+1(-1)$ denotes the retarded (advanced) component of $\widetilde{\mathcal{G}}$ and  $n(\varepsilon_{k_{1}},T)$ is the Fermi-Dirac occupation number (\ref{FermiDirac}).
The Fourier transformation with respect to the imaginary time \cite{Abrikosov1965} gives the mean-field Matsubara Green's function in the domain of the discrete
imaginary energy variable:
\begin{eqnarray}
\widetilde{\mathcal{G}}_{k_{2}k_{1}}(\varepsilon_{\ell}) = \frac{1}{2}\int^{1/T}_{-1/T}d\tau_{21}e^{i\varepsilon_{\ell}\tau_{21}}\widetilde{\mathcal{G}}(2,1)  
\nonumber \\
= \frac{\delta_{k_1k_2}}{i\varepsilon_{\ell}-\varepsilon_{k_{1}}+\mu} = \delta_{k_{1}k_{2}}\widetilde{\mathcal{G}}_{k_{1}}(\varepsilon_{\ell}).
\end{eqnarray} 
\begin{figure}
\includegraphics[scale=0.65]{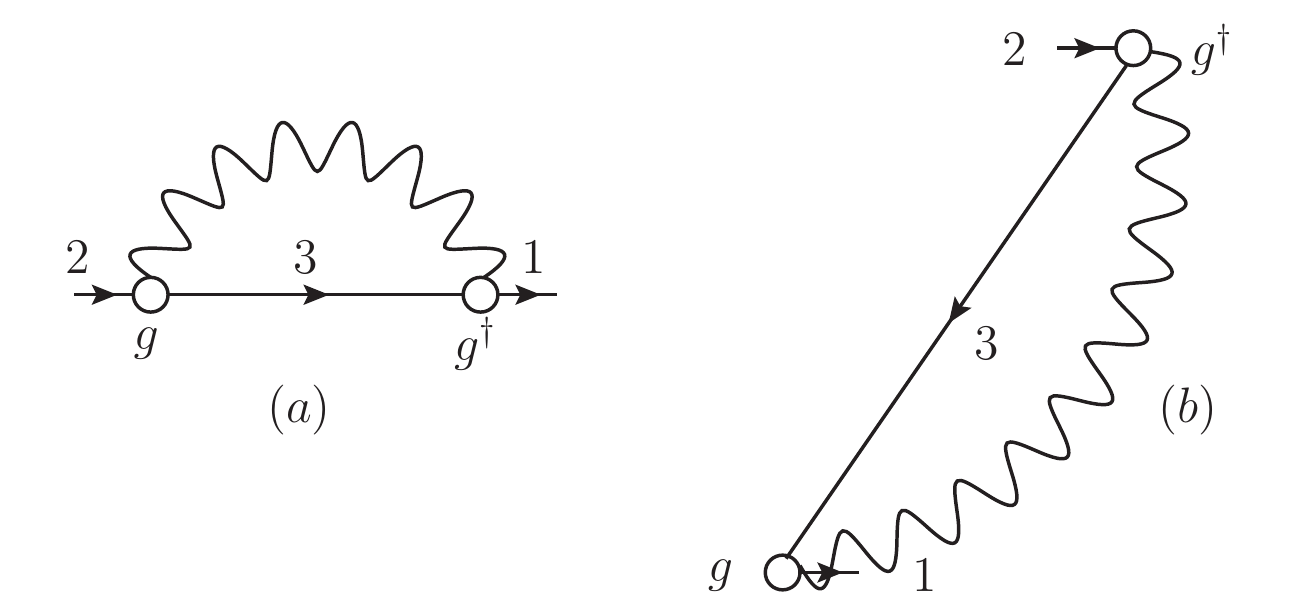}
\caption{Diagrammatic representation of $\Sigma^{e}$ for (a) $\sigma=+1$ and (b) $\sigma=-1$. Straight lines represent one-fermion propagators, circles denote the particle-phonon coupling vertices, and the wiggly lines correspond to the phonon propagators.}
\label{Diagrammatic of Sigma-e}
\end{figure}
As it is shown in the EOM approach \cite{VinhMau1979}, the Dyson equation for the single-particle Green's function with the exact self-energy can not be reduced to the closed form. The time-dependent part of the self-energy involves the knowledge about the three-body Green function, whose EOM can be further generated, but includes higher-order fermionic propagators. A factorization of the  three-body Green function into the one-body and two-body propagators allows for a truncation of the problem at the two-body level \cite{VinhMau1979,SchuckTohyama2016a}. 
Thus, the energy-dependent mass operator $\Sigma^{e}(\varepsilon)$ includes, in the leading approximation, the single-particle propagator contracted with the particle-hole and particle-particle (\textrm{pp}) response (correlation) functions, as shown diagrammatically in Fig. \ref{Diagrammatic of Sigma-e}. These correlated \textrm{ph} and \textrm{pp} pairs are identified with phonons propagating in the nuclear medium. The poles of the phonon propagators $\omega_m$, together with the particle-phonon, or particle-vibration, coupling vertices $g^{m}$ are, in general, extracted from the self-consistent solutions of the particle-hole and particle-particle response equations of the type (\ref{BSE}). In the leading approximation we can neglect the retardation effects in these internal correlation functions, i.e. take the solutions of the RPA type. This approximation will be considered in the present work. Further approximation neglects the contributions of pairing, charge-exchange and spin-flip vibrations, which are proven to be minor. The procedure of including higher-order effects was outlined in Ref. \cite{Litvinova2015} and will be considered numerically elsewhere.

The finite-temperature phonon propagator is defined as \cite{Abrikosov1965}:
\begin{equation}
\mathcal{D}_{m}(\omega_n)=\sum_{\sigma}\frac{\sigma}{i\omega_n-\sigma\omega_{m}},
\end{equation}
where $m$ labels the complete set of the phonon quantum numbers, in particular, $\omega_{m}$ is the real phonon frequency. The analytical form of the mass operator $\Sigma^{e}$ shown in Fig. \ref{Diagrammatic of Sigma-e} reads:
\begin{equation}
\Sigma^{e}_{k_{1}k_{2}}(\varepsilon_{\ell})=-T\sum_{k_{3},m}\sum_{\ell'}\widetilde{\mathcal{G}}_{k_{3}}(\varepsilon_{\ell'})\sum_{\sigma}\frac{\sigma g^{m(\sigma)\ast}_{k_{1}k_{3}}g_{k_{2}k_{3}}^{m(\sigma)}}{i\varepsilon_{\ell}-i\varepsilon_{\ell'}-\sigma\omega_{m}},
\label{Sigmae}
\end{equation}
where we define the vertices $g^{m(\sigma)}$ as \cite{Tselyaev1989,LitvinovaRingTselyaev2007}:
\begin{equation}
g^{m(\sigma)}_{k_{1}k_{2}}=\delta_{\sigma,+1}g^{m}_{k_{1}k_{2}}+\delta_{\sigma,-1}g_{k_{2}k_{1}}^{m\ast}.
\end{equation}
The phonon vertices $g_{k_{1}k_{2}}^{m}$ are, in the leading approximation,
\begin{equation}
g^{m}_{k_{1}k_{2}}=\sum_{k_{3}k_{4}}\widetilde{\mathcal{U}}_{k_{1}k_{2},k_{3}k_{4}}\rho^{m}_{k_{3}k_{4}},
\end{equation}
where 
\begin{equation}
\label{Static RMF Interaction}\widetilde{\mathcal{U}}_{k\ell,k'\ell'}=\frac{\delta\widetilde{\Sigma}_{k'\ell'}}{\delta\rho_{k\ell}}
\end{equation}
is the effective meson-exchange interaction in the static approximation and
$\rho^{m}$ are the transition densities of the phonons. 
The summation over $\ell'$ in Eq. (\ref{Sigmae})  can be transformed into a contour integral following the technique of Ref. \cite{Fetter2003}. Thus, the final expression for the mass operator $\Sigma^{e}$ reads:
\begin{eqnarray}
\Sigma^{e}_{k_{1}k_{2}}(\varepsilon_{\ell})&=&\sum_{k_{3},m}\Bigg\{g^{m\ast}_{k_{1}k_{3}}g^{m}_{k_{2}k_{3}}\frac{N(\omega_{m},T)+1-n(\varepsilon_{k_{3}},T)}{i\varepsilon_{\ell}-\varepsilon_{k_{3}}+\mu-\omega_{m}}+\nonumber\\
&+&g^{m}_{k_{3}k_{1}}g^{m\ast}_{k_{3}k_{2}}\frac{n(\varepsilon_{k_{3}},T)+N(\omega_{m},T)}{i\varepsilon_{\ell}-\varepsilon_{k_{3}}+\mu+\omega_{m}}\Bigg\},
\end{eqnarray}
where $N(\omega_{m},T)$ is the occupation number of $m$th phonon with frequency $\omega_{m}$:
\begin{equation}
N(\omega_{m},T)=\frac{1}{e^{\omega_{m}/T}-1}.
\end{equation}
Similar to the mass operator $\Sigma$, the interaction amplitude $\mathcal{U}$ has two terms, i.e., the static relativistic mean-field  part $\widetilde{\mathcal{U}}$ and energy-dependent part $\mathcal{U}^{e}$. The energy-dependent interaction amplitude $\mathcal{U}^{e}$, which satisfies the finite-temperature dynamical consistency condition, analogously to the $T=0$ case \cite{KamerdzhievTertychnyiTselyaev1997}, 
\begin{eqnarray}
\Sigma^{e}_{k_{1}k_{2}}(\varepsilon_{\ell}+\omega_{n})&-&\Sigma^{e}_{k_{1}k_{2}}(\varepsilon_{\ell})\nonumber\\
&=&T\sum_{k_{3}k_{4}}\sum_{\ell'}\mathcal{U}^{e}_{k_{2}k_{1},k_{4}k_{3}}(\omega_{n},\varepsilon_{\ell},\varepsilon_{\ell'})\times\nonumber\\
&\times&[\widetilde{\mathcal{G}}_{k_{3}k_{4}}(\varepsilon_{\ell'}+\omega_{n})-\widetilde{\mathcal{G}}_{k_{3}k_{4}}(\varepsilon_{\ell'})],
\end{eqnarray}
takes the form
\begin{eqnarray}
&\mathcal{U}^{e}_{k_{1}k_{2},k_{3}k_{4}}&(\omega_{n},\varepsilon_{\ell},\varepsilon_{\ell'})\nonumber\\
&=&\sum_{m}\frac{g^{m}_{k_{4}k_{2}}g^{m\ast}_{k_{3}k_{1}}}{i\varepsilon_{\ell}-i\varepsilon_{\ell'}+\omega_{m}}-\sum_{m}\frac{g^{m\ast}_{k_{2}k_{4}}g^{m}_{k_{1}k_{3}}}{i\varepsilon_{\ell}-i\varepsilon_{\ell'}-\omega_{m}}.\nonumber\\
\end{eqnarray}

\begin{figure*}
\includegraphics[scale=0.70]{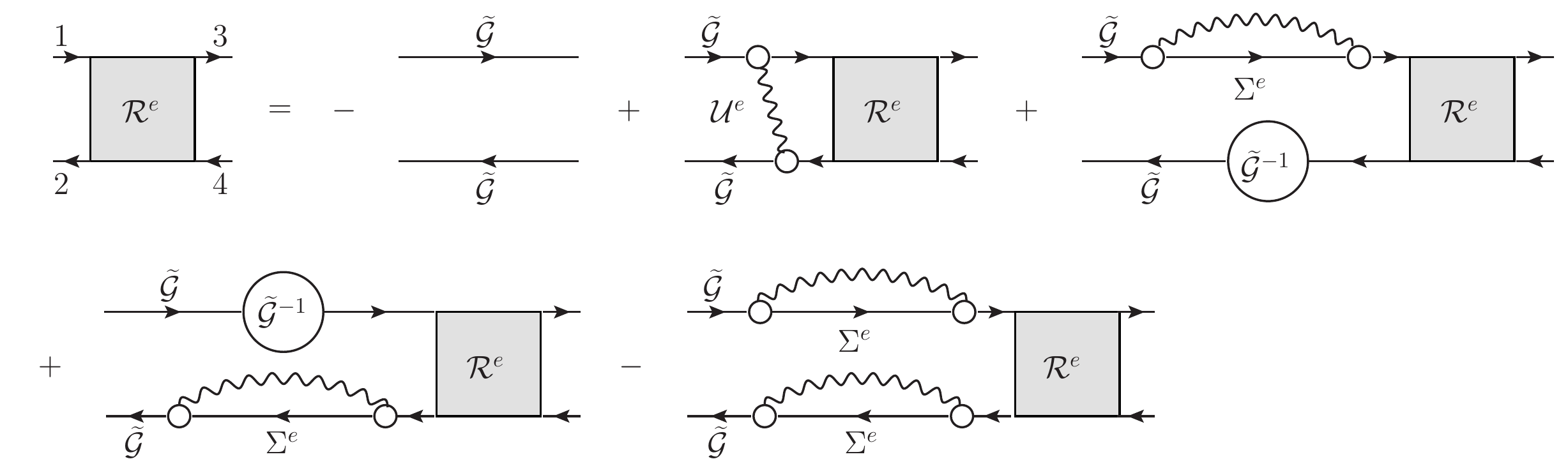}
\caption{Bethe-Salpeter equation for the correlated propagator $\mathcal{R}^{e}$ in the diagrammatic representation.}
\label{Diagrammatic of Re}
\end{figure*}

Similar to our treatment of the Dyson equation (\ref{Exact Dyson}), we can solve the BSE (\ref{BSE}) in two steps. First, we calculate the correlated propagator $\mathcal{R}^{e}$ from the BSE
\begin{equation}
\label{Operator Form of Re}\mathcal{R}^{e}=-\mathcal{G}\mathcal{G}+\mathcal{G}\mathcal{G}\mathcal{U}^{e}\mathcal{R}^{e}.
\end{equation}
Second, we solve the remaining equation
\begin{equation}
\mathcal{R}=\mathcal{R}^{e}-\mathcal{R}^{e}\widetilde{\mathcal{U}}\mathcal{R},
\label{FullResp}
\end{equation} 
to obtain the full response $\mathcal{R}$. 
In the approaches based on the well-defined mean field, such as the RMF, it is convenient to use the mean-field basis which diagonalizes the mean-field one-fermion Green's function $\widetilde{\mathcal{G}}$ defined by Eq. (\ref{Mean-Field Dyson}). Then, 
the equation for $\mathcal{R}^{e}$ can be formulated in terms of $\widetilde{\mathcal{G}}$: 
\begin{equation}
\label{Re in terms of Mean-Field G}\mathcal{R}^{e}=\widetilde{\mathcal{R}}^{0}-\widetilde{\mathcal{R}}^{0}\mathcal{W}^{e}\mathcal{R}^{e},
\end{equation}
where
\begin{eqnarray}
\label{Rtilde0}\widetilde{\mathcal{R}}^{0}&=&-\widetilde{\mathcal{G}}\widetilde{\mathcal{G}},\\
\label{mathcalWe}\mathcal{W}^{e}&=&W^{e}-\Sigma^{e}\Sigma^{e},\\
\label{We}W^{e}&=&\mathcal{U}^{e}+\widetilde{\mathcal{G}}^{-1}\Sigma^{e}+\Sigma^{e}\widetilde{\mathcal{G}}^{-1}.
\end{eqnarray}
The diagrammatic representation of Eq. (\ref{Re in terms of Mean-Field G}) is shown in Fig. (\ref{Diagrammatic of Re}).
\subsection{Finite-temperature time blocking approximation: "soft" blocking}
As the equation (\ref{Re in terms of Mean-Field G}) has  the singular kernel, it can not be solved directly in its present form.  The time blocking approximation proposed originally in Ref. \cite{Tselyaev1989} for the case of $T=0$ and adopted for the relativistic framework in Refs. \cite{LitvinovaRingTselyaev2007,LitvinovaRingTselyaev2008} allows for a reduction of the BSE to one energy variable equation with the interaction kernel where the internal energy variables can be integrated out separately. The main idea of the method is to introduce a time projection operator into the integral part of the BSE for the correlated propagator $\mathcal{R}^{e}$. This operator acting on the uncorrelated mean-field propagator in the second term on the right hand side of Eq. (\ref{Re in terms of Mean-Field G}) brings it to a separable form with respect to its two energy variables, see Ref. \cite{KamerdzhievTertychnyiTselyaev1997}  for details.
The analogous imaginary-time projection operator for the finite temperature case would look as follows:
\begin{eqnarray}
\Theta(12,34)&=&\delta_{\sigma_{k_{1}},-\sigma_{k_{2}}}\theta(\sigma_{k_{1}}\tau_{41})\theta(\sigma_{k_{1}}\tau_{32}),
\label{t-proj}
\end{eqnarray}
however, it turns out that at $T>0$ it does not lead to a similar separable function in the kernel of Eq. (\ref{Re in terms of Mean-Field G}).  

In order to reach the desired separable form, we found that the 
imaginary-time projection operator has to be modified as follows: 
\begin{eqnarray}
\Theta(12,34;T)=\delta_{\sigma_{k_{1}},-\sigma_{k_{2}}}\theta(\sigma_{k_{1}}\tau_{41})\theta(\sigma_{k_{1}}\tau_{32})\times\nonumber\\
\times[n(\sigma_{k_{1}}\varepsilon_{k_{2}},T)\theta(\sigma_{k_{1}}\tau_{12})
+n(\sigma_{k_{2}}\varepsilon_{k_{1}},T)\theta(\sigma_{k_{2}}\tau_{12})],\nonumber\\
\label{t-proj-ft}
\end{eqnarray}
i.e. it should contain an additional multiplier with the dependence on the diffuse Fermi-Dirac distribution function, which turns to unity in the $T=0$ limit at the condition $\sigma_{k_{1}} =-\sigma_{k_{2}}$.

Acting by the projection operator $\Theta(12,34;T)$ on the components of $\widetilde{\mathcal{R}}^{0}(12,34)=-\widetilde{\mathcal{G}}(3,1)\widetilde{\mathcal{G}}(2,4)$, we construct an operator $\widetilde{\mathcal{D}}$ of the form
\begin{eqnarray}
\label{Operator D}
\widetilde{\mathcal{D}}(12,34) = \Theta(12,34;T)\widetilde{\mathcal{R}}^{0(\sigma_{k_{1}}\sigma_{k_{2}})}(12,34) =
\nonumber\\
=-\delta_{\sigma_{k_{1}},-\sigma_{k_{2}}}\widetilde{\mathcal{G}}^{\sigma_{k_{1}}}(3,1)\widetilde{\mathcal{G}}^{\sigma_{k_{2}}}(2,4)\theta(\sigma_{k_{1}}\tau_{41})\theta(\sigma_{k_{1}}\tau_{32})\times\nonumber\\
\times[n(\sigma_{k_{1}}\varepsilon_{k_{2}},T)\theta(\sigma_{k_{1}}\tau_{12})
+
n(\sigma_{k_{2}}\varepsilon_{k_{1}},T)\theta(\sigma_{k_{2}}\tau_{12})]\nonumber \\
\end{eqnarray}
and make a substitution
\begin{equation}
\widetilde{\mathcal{R}}^{0}\rightarrow \widetilde{\mathcal{D}}
\end{equation}
in the second term of Eq. (\ref{Re in terms of Mean-Field G}). In Eq. (\ref{Operator D}), $\sigma_{k}=+1(-1)$ for particle (hole). The Kronecker delta $\delta_{\sigma_{k_{1}},-\sigma_{k_{2}}}$ constraints the possible combinations of $(\sigma_{k_{1}},\sigma_{k_{2}})$ to be ($+1,-1$) and ($-1,+1$). A pair of state $\{k_{1},k_{2}\}$ is considered as a \textrm{ph} (\textrm{hp}) pair if the energy difference $\varepsilon_{k_{1}}-\varepsilon_{k_{2}}$ is larger (smaller) than zero. The replacement of $\widetilde{\mathcal{R}}^{0}$ by $\widetilde{\mathcal{D}}$ corresponds to the elimination of the processes with configuration more complex than \textrm{ph}$\otimes$phonon ones, in analogy to the case of $T=0$ \cite{KamerdzhievTertychnyiTselyaev1997}. Thus, in the leading approximation, we keep the terms with \textrm{ph} and \textrm{ph}$\otimes$phonon configurations and neglect terms with higher configurations including $-\Sigma^{e}\Sigma^{e}$ in Eq. (\ref{Re in terms of Mean-Field G}). As a result, the equation for the correlated propagator ${\mathcal{R}}^{e}$ takes the form
\begin{eqnarray}
\label{BSE for Re Tilde}&{\mathcal{R}}^{e}&(12,34)=\widetilde{\mathcal{R}}^{0}(12,34)+\sum_{5678}^{\tau}\widetilde{\mathcal{D}}(12,56)\Big[\mathcal{U}^{e}(56,78)+\nonumber\\
&+&\widetilde{\mathcal{G}}^{-1}(7,5)\Sigma^{e}(6,8)+\Sigma^{e}(7,5)\widetilde{\mathcal{G}}^{-1}(6,8)\Big]{\mathcal{R}}^{e}(78,34),\nonumber\\
\end{eqnarray}
in the imaginary-time representation, where the summations imply integrations over the time arguments:
\begin{equation}
\sum^{\tau}_{12..}=\sum_{k_{1}k_{2}...}\int_{0}^{1/T}d\tau_{1}d\tau_{2}\cdots.
\end{equation}

After the 3-Fourier transformation with respect to the imaginary-time,
\begin{eqnarray}
&{\mathcal{R}}^{e}_{k_{1}k_{2},k_{3}k_{4}}&(\omega_{n},\varepsilon_{\ell},\varepsilon_{\ell'})=\frac{1}{8}\int_{-1/T}^{1/T}d\tau_{31}d\tau_{21}d\tau_{34}\times\nonumber\\
&&\times e^{i(\omega_{n}\tau_{31}+\varepsilon_{\ell}\tau_{21}+\varepsilon_{\ell'}\tau_{34})}{\mathcal{R}}^{e}(12,34),
\end{eqnarray}
the summation over the fermionic discrete variables $\ell$ and $\ell'$, 
\begin{eqnarray}
{\mathcal{R}}^{e}_{k_{1}k_{2},k_{3}k_{4}}(\omega_{n})=T^{2}\sum_{\ell}\sum_{\ell'}{\mathcal{R}}^{e}_{k_{1}k_{2},k_{3}k_{4}}(\omega_{n},\varepsilon_{\ell},\varepsilon_{\ell'}),\nonumber\\
\end{eqnarray}
and the analytical continuation to the real frequencies, we obtain 
\begin{eqnarray}
\label{BSE for Correlated Propagator Energy Reps}&&{\mathcal{R}}^{e}_{k_{1}k_{2},k_{3}k_{4}}(\omega)=\widetilde{\mathcal{R}}^{0}_{k_{1}k_{2},k_{3}k_{4}}(\omega)-\nonumber\\
&&-\sum_{k_{5}k_{6},k_{7}k_{8}}\widetilde{\mathcal{R}}^{0}_{k_{1}k_{2},k_{5}k_{6}}(\omega){\Phi}_{k_{5}k_{6},k_{7}k_{8}}(\omega){\mathcal{R}}^{e}_{k_{7}k_{8},k_{3}k_{4}}(\omega),\nonumber\\
\end{eqnarray}
where the spectral representation of the uncorrelated propagator is 
\begin{eqnarray}
\label{Free Response}\widetilde{\mathcal{R}}^{0}_{k_{1}k_{2},k_{3}k_{4}}(\omega)&=&-\delta_{k_{1}k_{3}}\delta_{k_{2}k_{4}}\frac{n(\varepsilon_{k_{2}},T)-n(\varepsilon_{k_{1}},T)}{\omega-\varepsilon_{k_{1}}+\varepsilon_{k_{2}}}\nonumber\\
\end{eqnarray}
and the particle-vibration coupling amplitude is 
\begin{eqnarray}
\label{Particle-Vibration Coupling Amplitude}{\Phi}_{k_{1}k_{2},k_{3}k_{4}}(\omega)=\frac{\delta_{\sigma_{k_{1}},-\sigma_{k_{2}}}\sigma_{k_{1}}}{n(\varepsilon_{k_{4}},T)-n(\varepsilon_{k_{3}},T)}\times\nonumber\\
\times\sum_{k_{5}k_{6}m}\sum_{\eta_{m}=\pm 1}\eta_{m}\zeta^{m\eta_{m}}_{k_{1}k_{2},k_{5}k_{6}}\zeta^{m\eta_{m}\ast}_{k_{3}k_{4},k_{5}k_{6}}\times\nonumber\\
\times\frac{[N(\eta_{m}\omega_{m},T)+n(\varepsilon_{k_{6}},T)]
[n(\varepsilon_{k_{6}}-\eta_{m}\omega_{m},T)-n(\varepsilon_{k_{5}},T)]}{\omega-\varepsilon_{k_{5}}+\varepsilon_{k_{6}}-\eta_{m}\omega_{m}},\nonumber\\
\end{eqnarray}
where we have defined the phonon vertex matrices $\zeta^{m\eta_{m}}$ as
\begin{eqnarray}
\zeta^{m\eta_{m}}_{k_{1}k_{2},k_{5}k_{6}}&=&\delta_{k_{1}k_{5}}g^{m(\eta_{m})}_{k_{6}k_{2}}-g^{m(\eta_{m})}_{k_{1}k_{5}}\delta_{k_{6}k_{2}}.
\end{eqnarray}
It can be verified that the \textrm{hp}-components of particle-vibration coupling amplitude, i.e., ${\Phi}_{\textrm{hp,h'p'}}(\omega)$, are connected to the \textrm{ph}-components, i.e., ${\Phi}_{\textrm{ph,p'h'}}(\omega)$, via
\begin{equation}
{\Phi}_{\textrm{hp,h'p'}}(\omega)={\Phi}^{\ast}_{\textrm{ph,p'h'}}(-\omega).
\end{equation}
Using the shorthand notation, Eq. (\ref{BSE for Correlated Propagator Energy Reps}) can be rewritten as
\begin{equation}
{\mathcal{R}}^{e}=\widetilde{\mathcal{R}}^{0}-\widetilde{\mathcal{R}}^{0}{\Phi}{\mathcal{R}}^{e}
\end{equation}
and the BSE for the full response function becomes
\begin{equation}
\label{BSE Full Response with Correction}\mathcal{R}=\widetilde{\mathcal{R}}^{0}-\widetilde{\mathcal{R}}^{0}\left[\widetilde{\mathcal{U}}+\delta{\Phi}\right]\mathcal{R},
\end{equation}
where the particle-phonon coupling amplitude is corrected as follows 
\begin{equation}
{\Phi}(\omega)\rightarrow\delta{\Phi}(\omega)={\Phi}(\omega)-{\Phi}(0),
\end{equation}
i.e. by the subtraction of itself at $\omega=0$. This subtraction is necessary for the CEDF-based calculations where the static contribution of the particle-phonon coupling is implicitly contained in the residual interaction $\widetilde{\mathcal{U}}$ \cite{Tselyaev2013,LitvinovaRingTselyaev2007}.
%
\subsection{Strength function and transition densities}
%
In order to obtain the strength distribution 
we, in fact, do not necessarily need to know the response function, because the strength distribution is directly related to the nuclear polarizability (\ref{polarizability}). Thus, instead of solving the equation (\ref{BSE Full Response with Correction}), we can consider a single or a double convolution of it with the external field. In particular, it is often useful to consider the density matrix variation $\delta\rho$:
\begin{eqnarray}
\delta\rho_{k_{1}k_{2}}(\omega)&=&-\sum_{k_{3}k_{4}}\mathcal{R}_{k_{1}k_{2},k_{3}k_{4}}(\omega)V^{0}_{k_{4}k_{3}},\\
\delta\rho_{k_{1}k_{2}}^{0}(\omega)&=&-\sum_{k_{3}k_{4}}\widetilde{\mathcal{R}}^{0}_{k_{1}k_{2},k_{3}k_{4}}(\omega)V^{0}_{k_{4}k_{3}},
\end{eqnarray}
so that  Eq. (\ref{BSE Full Response with Correction}) can be transformed into
\begin{equation}
\label{Equation for Density Matrix Variation}\delta\rho(\omega)=\delta\rho^{0}(\omega)-\widetilde{\mathcal{R}}^{0}(\omega)\left[\widetilde{\mathcal{U}}+\delta{\Phi}(\omega)\right]\delta\rho(\omega). 
\end{equation}
Thus, the spectral density ${S}(E)$ can be expressed as 
\begin{equation}
\label{Strength Function in term of Density Matrix Variation}{S}(E)=-\frac{1}{\pi}\lim_{\Delta\rightarrow+0}\text{Im}\sum_{k_{1}k_{2}}V^{0\ast}_{k_{2}k_{1}}\delta\rho_{k_{1}k_{2}}(E+i\Delta).
\end{equation}
The transition density
\begin{equation}
\rho^{fi}_{k_{1}k_{2}}=\langle f|\hat{a}_{k_{1}}^{\dag}\hat{a}_{k_{2}}|i\rangle
\end{equation}
from the initial $|i\rangle$ to the final $|f\rangle$ state  
can be then related to the spectral density ${S}(E)$ at the energy $E=\omega_{fi}$. In the vicinity of $\omega_{fi}$ the full response function is a simple pole: 
\begin{equation}
\label{Simple Pole Structure}\left.\mathcal{R}^{fi}_{k_{1}k_{2},k_{3}k_{4}}(\omega)\right|_{\omega\approx\omega_{fi}}\approx-\frac{\rho^{fi}_{k_{1}k_{2}}\rho^{fi\ast}_{k_{3}k_{4}}}{\omega-\omega_{fi}},
\end{equation}
so that the imaginary part of the matrix element $\delta\rho_{k_{1}k_{2}}(\omega_{fi}+i\Delta)$ takes the form:
\begin{equation}
\text{Im}\delta\rho_{k_{1}k_{2}}(\omega_{fi}+i\Delta)=-\frac{1}{\Delta}\rho^{fi}_{k_{1}k_{2}}\sum_{k_{3}k_{4}}\rho^{fi\ast}_{k_{3}k_{4}}V^{0}_{k_{4}k_{3}},
\end{equation}  
and the spectral density ${S}(\omega_{fi})$ is given by
\begin{equation}
{S}(\omega_{fi})=\frac{1}{\pi}\lim_{\Delta\rightarrow+0}\frac{1}{\Delta}\left|\sum_{k_{3}k_{4}}\rho^{fi\ast}_{k_{3}k_{4}}V^{0}_{k_{4}k_{3}}\right|^{2}.
\end{equation}
Combining the last two equations, we obtain, in analogy to the $T=0$ case \cite{LitvinovaRingTselyaev2007}, the relation:
\begin{equation}
\label{Transition Density in term of S(E) and Delta Rho}\rho_{k_{1}k_{2}}^{fi}=\lim_{\Delta\rightarrow+0}\sqrt{\frac{\Delta}{\pi {S}(\omega_{fi})}}\text{Im}\delta\rho_{k_{1}k_{2}}(\omega_{fi}+i\Delta),
\end{equation}
which allows for an extraction of the transition densities from a continuous strength distribution.  
To derive the normalization of the transition densities, it is convenient to rewrite Eq. (\ref{BSE Full Response with Correction}) in the form:
\begin{equation}
\label{Modified BSE}\left[(\widetilde{\mathcal{R}}^{0})^{-1}+\widetilde{\mathcal{U}}+{\Phi}(\omega)-{\Phi}(0)\right]\mathcal{R}(\omega)=1.
\end{equation} 
Taking the derivative of Eq. (\ref{Modified BSE}) with respect to $\omega$ gives
\begin{equation}
\label{Derivative of BSE}-\frac{d\mathcal{R}(\omega)}{d\omega}=\mathcal{R}(\omega)\frac{d(\widetilde{\mathcal{R}}^{0})^{-1}}{d\omega}\mathcal{R}(\omega)+\mathcal{R}(\omega)\frac{d{\Phi}(\omega)}{d\omega}\mathcal{R}(\omega).
\end{equation}
Inserting Eqs. (\ref{Free Response}) and (\ref{Simple Pole Structure}) into Eq. (\ref{Derivative of BSE}), we obtain the generalized normalization condition
\begin{equation}
\label{Generalized Normalization Condition}\rho^{fi\ast}\left[N-\left.\frac{d{\Phi}(\omega)}{d\omega}\right|_{\omega=\omega_{fi}}\right]\rho^{fi}=1,
\end{equation}
where $N$ is the finite-temperature RPA (FT-RPA) norm:
\begin{equation}
N_{k_{1}k_{2},k_{3}k_{4}}=\frac{\delta_{k_{1}k_{3}}\delta_{k_{2}k_{4}}}{n(\varepsilon_{k_{2}},T)-n(\varepsilon_{k_{1}},T)}.
\end{equation}
For the case of the energy-independent interaction, when the derivative of ${\Phi}(\omega)$ with respect to $\omega$ vanishes, we obtain the FT-RPA normalization:
\begin{equation}
\sum_{\textrm{ph}}\frac{|\rho^{fi}_{\textrm{ph}}|^{2}-|\rho^{fi}_{\textrm{hp}}|^{2}}{n(\varepsilon_{\textrm{h}},T)-n(\varepsilon_{\textrm{p}},T)}=1.
\end{equation}

\section{NUMERICAL DETAILS, RESULTS, AND DISCUSSION}
\subsection{Numerical details}
In this Section, we apply the finite-temperature relativistic time-blocking approximation developed above to a quantitative description of the IV GDR in the even-even spherical nuclei $^{48}$Ca, $^{68}\text{Ni}$ and $^{100,120,132}\text{Sn}$. The general scheme of the calculations is as follows: 
\begin{enumerate}
\item[(i)] The closed set of the RMF equations, i.e., Eqs. (\ref{Basis for Dirac Hamiltonian}), (\ref{RMF Equation for Sigma Meson})-(\ref{RMF Equation for Photon}) with the densities of Eqs. (\ref{FT-Scalar})-(\ref{FT-Charge}), are solved simultaneously in a self-consistent way using the  NL3 parameter set \cite{Lalazissis1997a} of the non-linear sigma-model. Thus, we obtain the temperature-dependent single-particle basis in terms of the Dirac spinors and the corresponding single-nucleon energies (\ref{Basis for Dirac Hamiltonian}). 
\item[(ii)] Using the obtained single-particle basis, the FT-RRPA equations, which are equivalent to Eq. (\ref{Equation for Density Matrix Variation}) without the particle-phonon coupling amplitude ${\Phi}(\omega)$, are solved with the static RMF residual interaction $\tilde{\mathcal{U}}$ of Eq. \eqref{Static RMF Interaction}, to obtain the phonon vertices $g^{m}$ and frequencies $\omega_{m}$. 
The set of phonons, 
together with the single-particle basis, forms the \textrm{1p1h}$\otimes$phonon configurations for the particle-phonon coupling amplitude ${\Phi}(\omega)$.
\item[(iii)] Finally, we solve Eq. (\ref{Equation for Density Matrix Variation}) and compute the strength function 
according to Eq. 
(\ref{Strength}) with the external field of the electromagnetic dipole character:
\begin{equation}
V^{0}_{1M} = \frac{eN}{A}\sum\limits_{i=1}^Z r_iY_{1M}(\Omega_i) - \frac{eZ}{A}\sum\limits_{i=1}^N r_iY_{1M}(\Omega_i) 
\end{equation} 
corrected for the center of mass motion. An alternative numerical solution in the momentum space is also implemented. In this case, we first solve Eq. (\ref{BSE for Correlated Propagator Energy Reps}) with $\Phi(\omega) \to \delta\Phi(\omega) $ in the basis of Dirac spinors and then Eq. (\ref{FullResp}) in the momentum-channel representation \cite{LitvinovaRingTselyaev2008}. Coincidence of the two solutions is used for the testing purposes. The momentum-channel representation has the advantage of a faster execution for large masses and high temperatures, while the Dirac-space representation allows for a direct extraction of the transition densities.
\end{enumerate}

In both representations, the amplitude $\Phi(\omega)$ takes the non-zero values only in a 25-30 MeV window with respect to the particle-hole energy differences. It has been verified by direct calculations that further extension of this window does not change noticeably the results for the strength distributions at the energies below this value.
The particle-hole basis was fixed by the limits $\varepsilon_{\textrm{ph}}\leq100$ MeV and $\varepsilon_{\alpha \textrm{h}}\geq-1800$ MeV with respect to the positive continuum. We have, however, directly verified that calculations with $\varepsilon_{\textrm{ph}}\leq300$ MeV eliminate the spurious translational mode completely, but do almost not change the physical states of the excitation spectra. The values of the smearing parameter 500 keV and 200 keV were adopted for the calculations of the medium-light and medium-heavy nuclei, respectively. The collective vibrations with quantum numbers of spin and parity $J^{\pi}=2^{+},\;3^{-},\;4^{+},\;5^{-},\;6^{+}$ and below 15 MeV were included in the phonon space. The phonon space was additionally truncated according to the values of the reduced transition probabilities of the corresponding electromagnetic transitions:  all modes with the values of the reduced transition probabilities less than 5\% of the maximal one were neglected. Keeping the last criterion, in particular, lead to a very strong increase of the number of phonons included in the model space at high temperatures: at $T\approx 5-6$ MeV this number becomes an order of magnitude larger than at $T=0$. This is related to the fact that at finite temperatures the typical collective modes lose their collectivity and many non-collective modes appear due to the thermal unblocking.

\subsection{Thermal mean field calculations for compound nuclei}

\begin{figure} 
\includegraphics[scale=0.35]{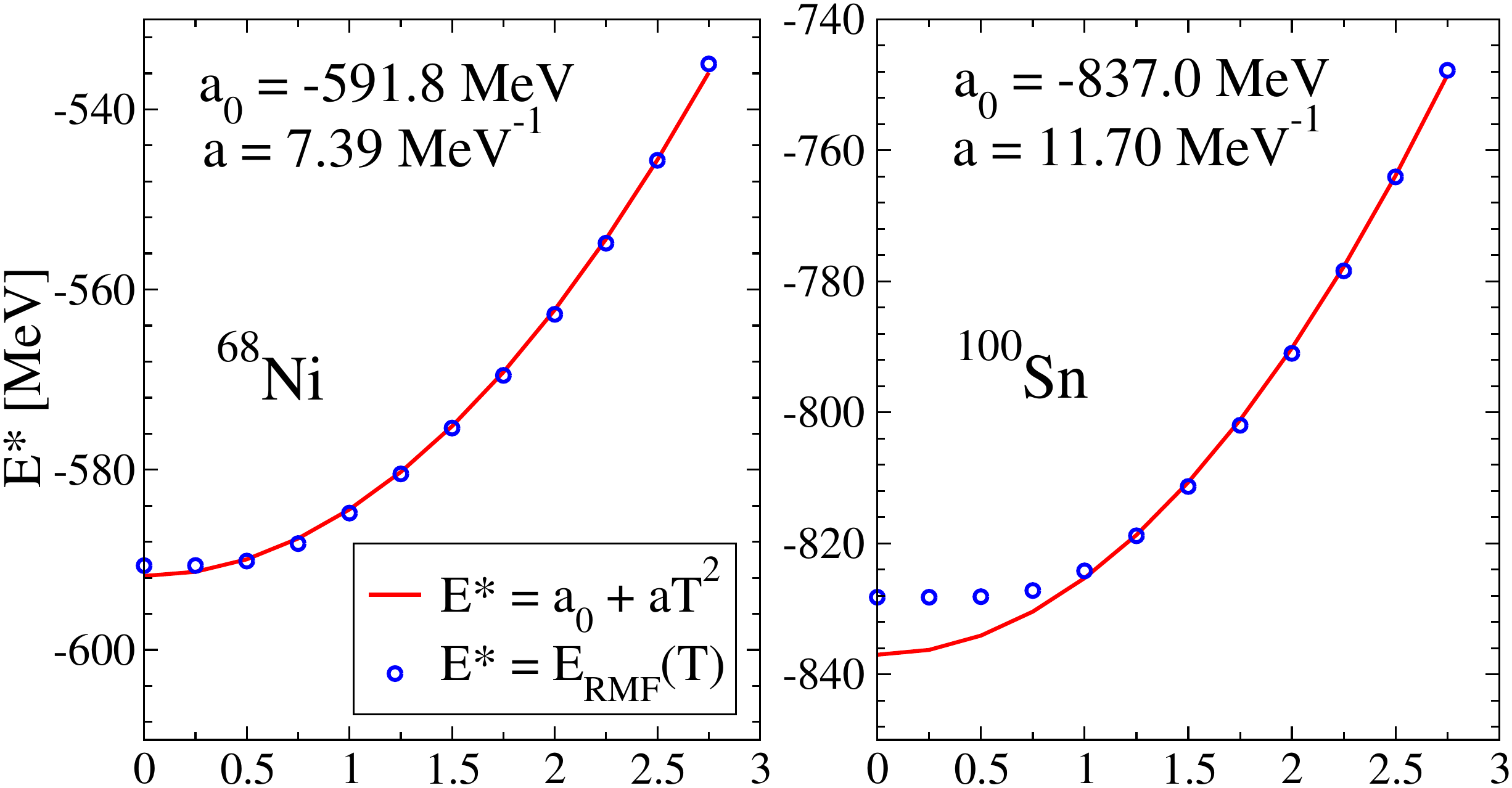}
\includegraphics[scale=0.35]{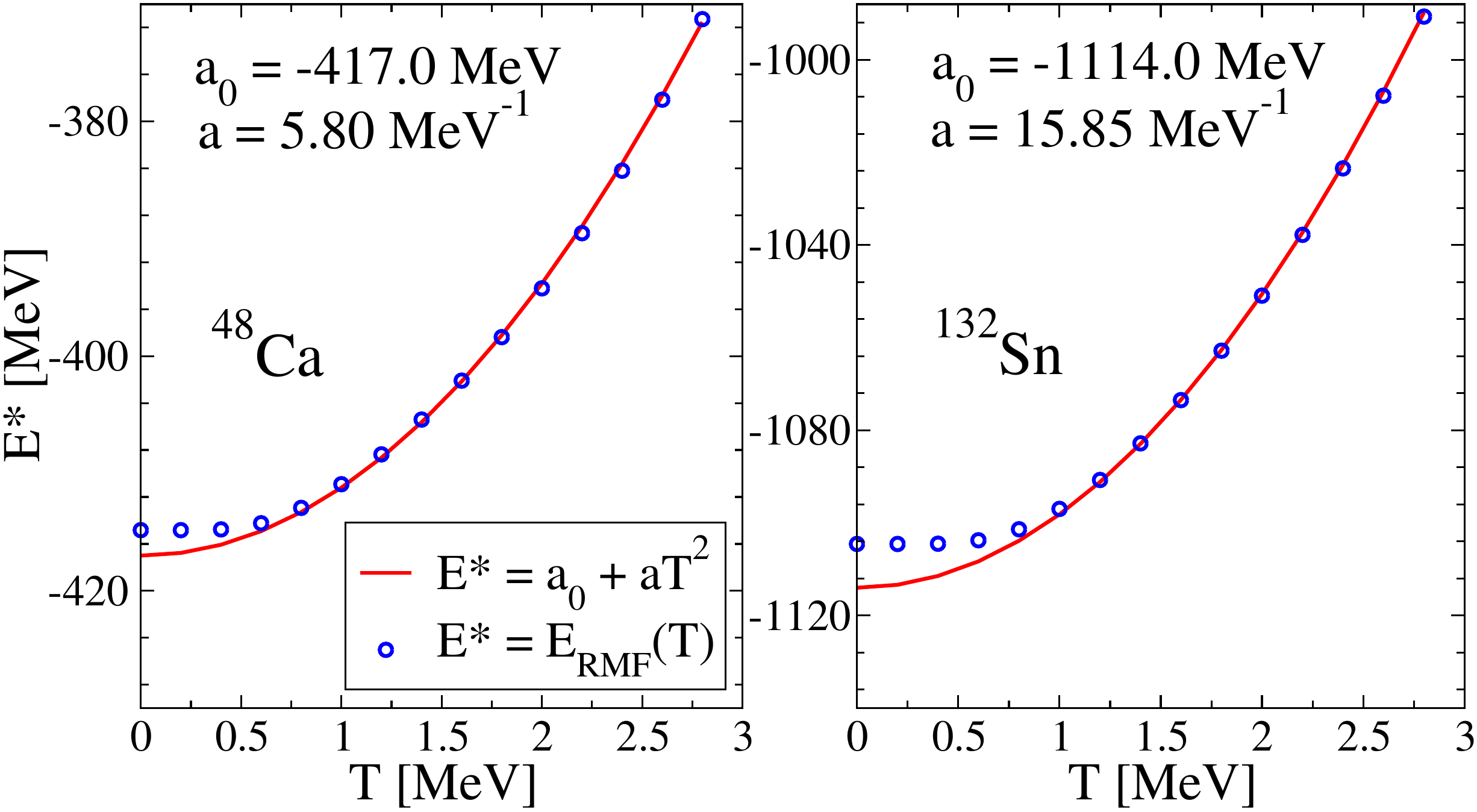}
\caption{The energies of the thermally excited nuclei $^{48}$Ca, $^{68}\text{Ni}$ and $^{100,132}\text{Sn}$ as functions of temperature: RMF (blue circles) and parabolic fits (red curves).}
\label{Level Density Parameter}
\end{figure}

The thermal RMF calculations of the excitation energy $E^{\ast}$ as a function of temperature $T$ for compound nuclei $^{48}$Ca, $^{68}\text{Ni}$ and $^{100,132}\text{Sn}$ are illustrated in Fig. \ref{Level Density Parameter}. Technically, as it follows from Section \ref{FT-RMF}, the effect of finite temperature on the total energy of a thermally excited nucleus is mainly induced by the change of the fermionic occupation numbers from the values of zero and one at $T=0$ to the Fermi-Dirac distribution (\ref{FermiDirac}). The fermionic densities of Eqs. (\ref{FT-Scalar}-\ref{FT-Charge}) change accordingly and, thus, affect the meson and photon fields being the sources for Eqs. (\ref{RMF Equation for Sigma Meson}-\ref{RMF Equation for Photon}). In turn, the changed meson fields give the feedback on the nucleons, so that the thermodynamical equilibrium is achieved through the self-consistent set of the thermal RMF equations. As the nucleons start to be promoted to higher-energy orbits with the temperature increase, the total energy should grow continuously and, in  principle, the dependence $E^{\ast}(T)$ has to be parabolic, in accordance with the non-interacting Fermi gas behavior. However, the discrete shell structure and especially the presence of the large shell gaps right above the Fermi surface in the doubly-magic nuclei cause a flat behavior of the excitation energy until the temperature values become sufficient to promote the nucleons over the shell gaps. This effect is clearly visible in Fig. \ref{Level Density Parameter} for the doubly-magic nuclei $^{48}$Ca and $^{100,132}\text{Sn}$, while it is much smaller in $^{68}\text{Ni}$ which has an open shell in the neutron subsystem. Otherwise, at $T\geq$ 1 MeV the thermal RMF $E^{\ast}(T)$ dependencies can be very well approximated by the parabolic fits providing the level density parameters which are close to the empirical Fermi gas values $a = A/k$, where $8<k<12$.

\subsection{Isovector Dipole Resonance in $^{48}$Ca, $^{68}\text{Ni}$ and $^{100,120,132}\text{Sn}$}
\begin{figure}
\hspace{-2mm}\includegraphics[scale=0.35]{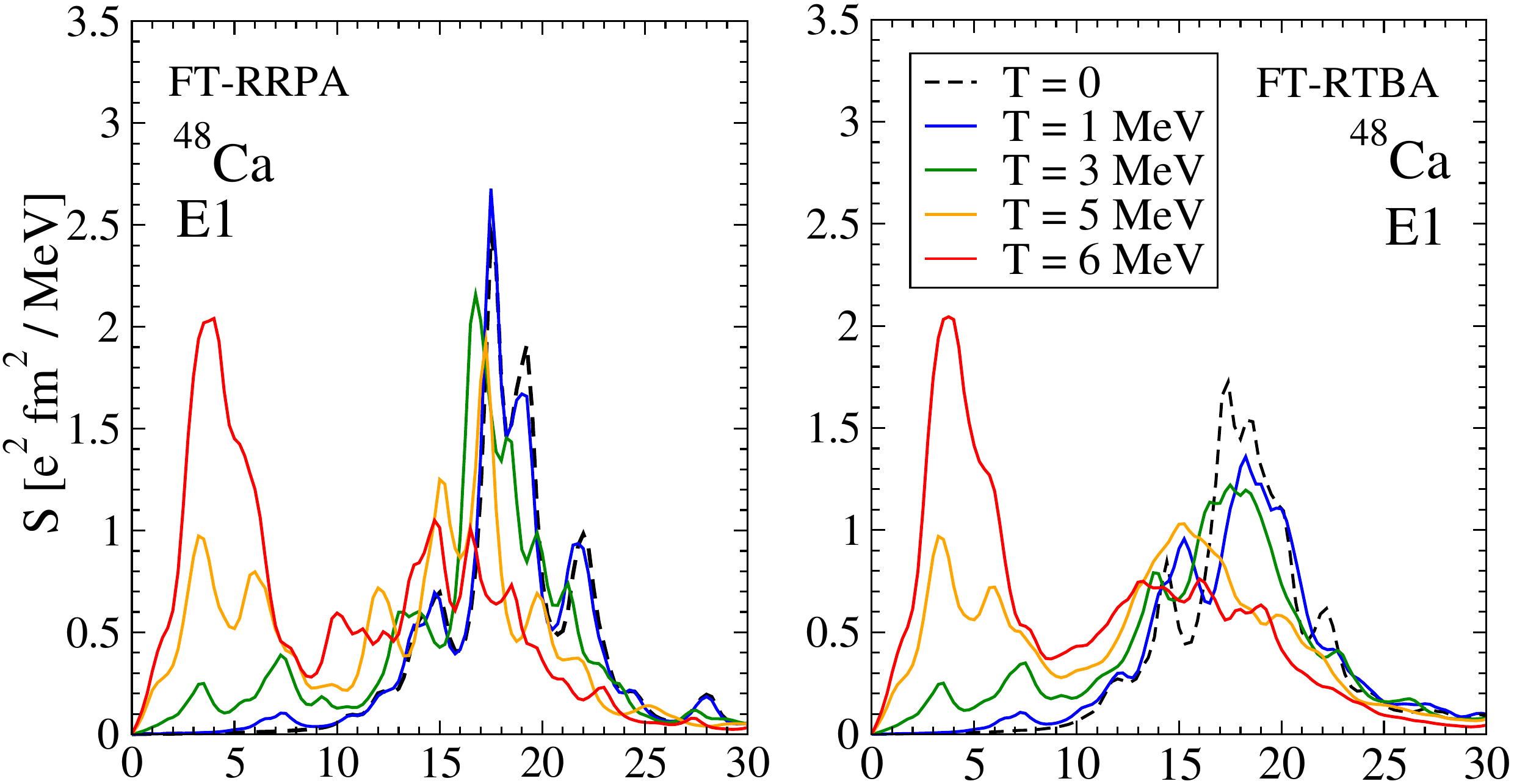}
\includegraphics[scale=0.35]{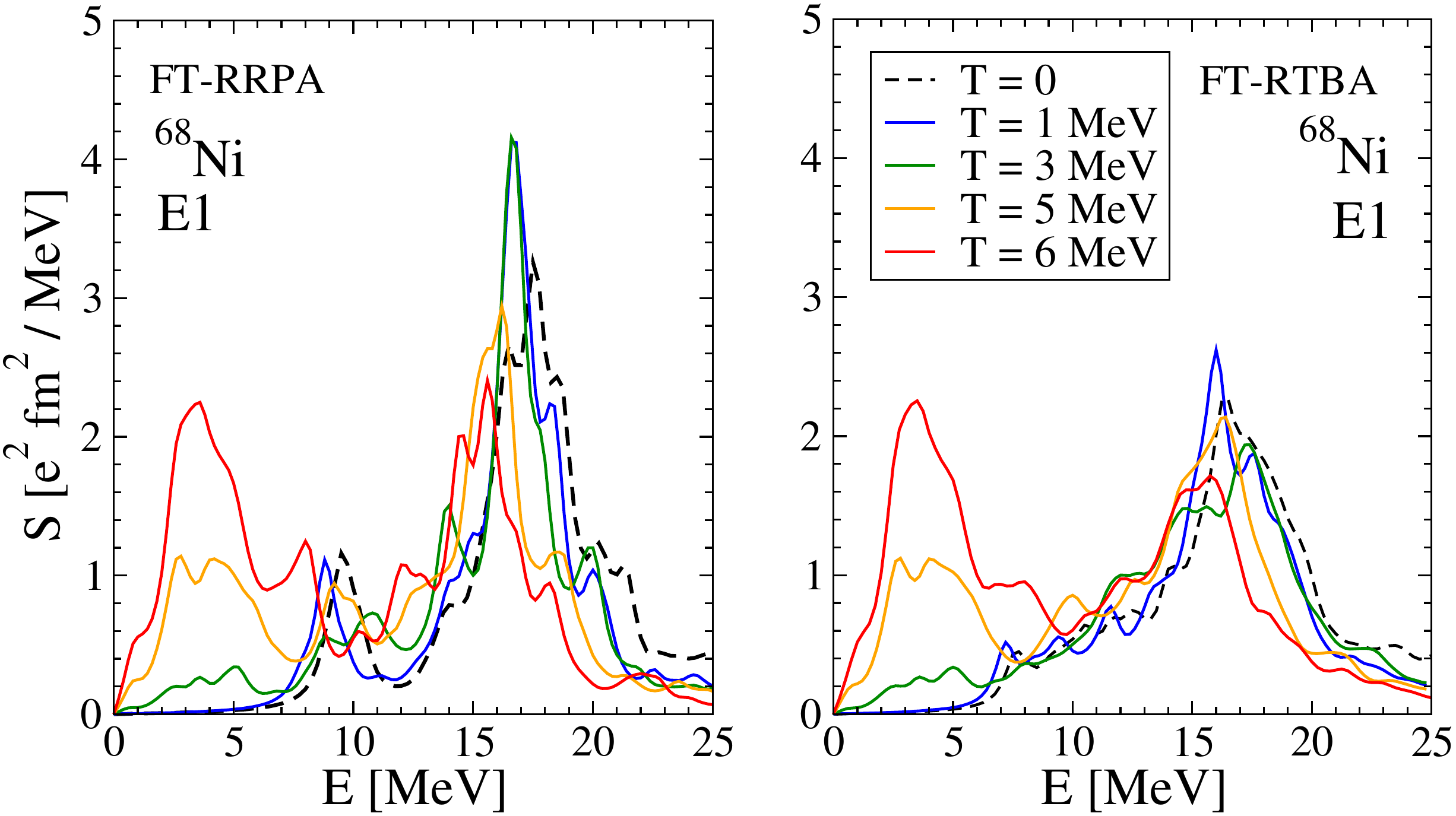}
\caption{Electric dipole spectral density in $^{48}$Ca and $^{68}\text{Ni}$ nuclei calculated within FT-RRPA (left panels) and FT-RTBA (right panels) at various temperatures. The value of smearing parameter $\Delta=500$ keV was adopted in both calculations.}
\label{CaNi}
\end{figure}
\begin{figure}
\includegraphics[scale=0.35]{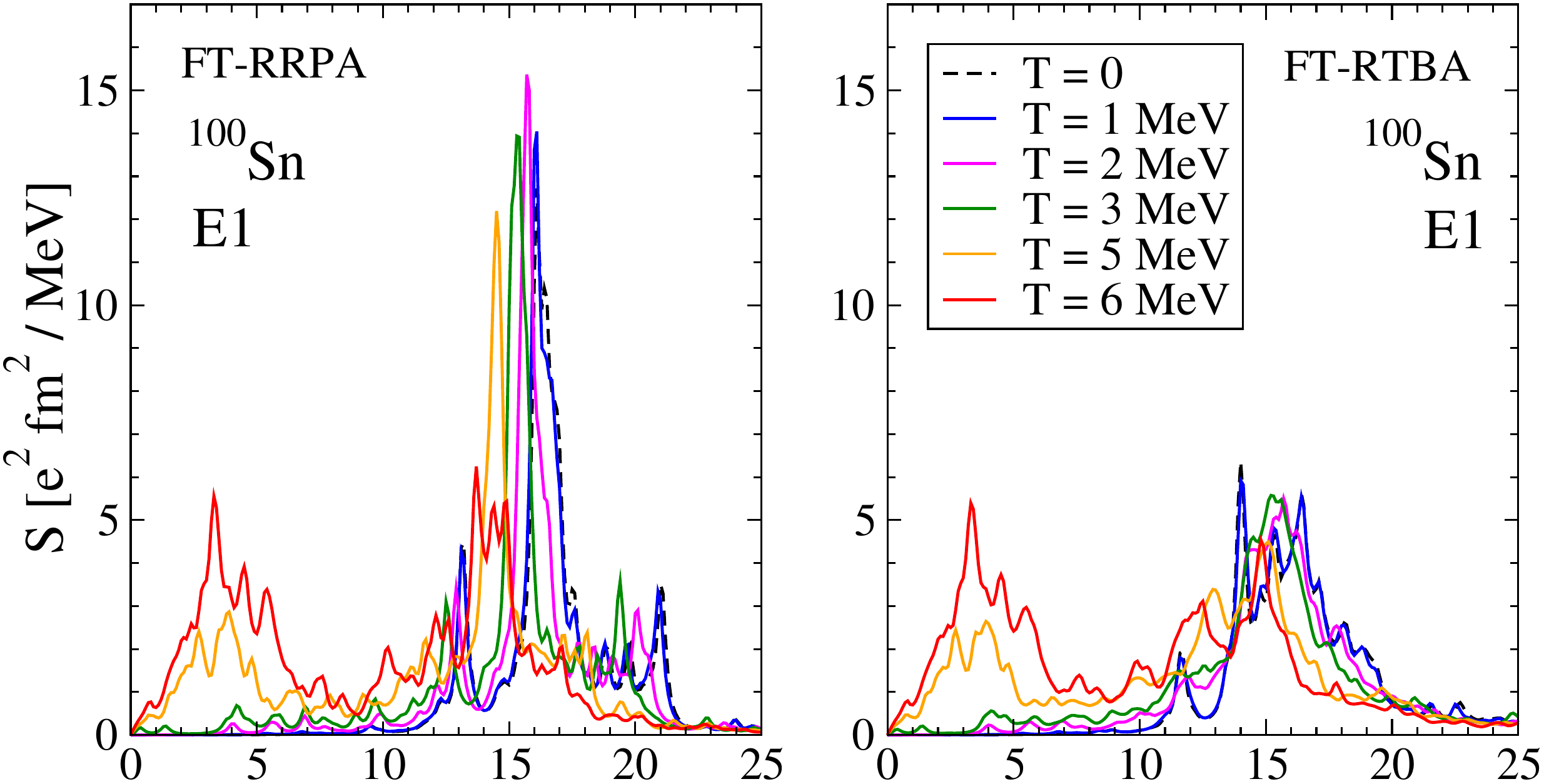}
\includegraphics[scale=0.35]{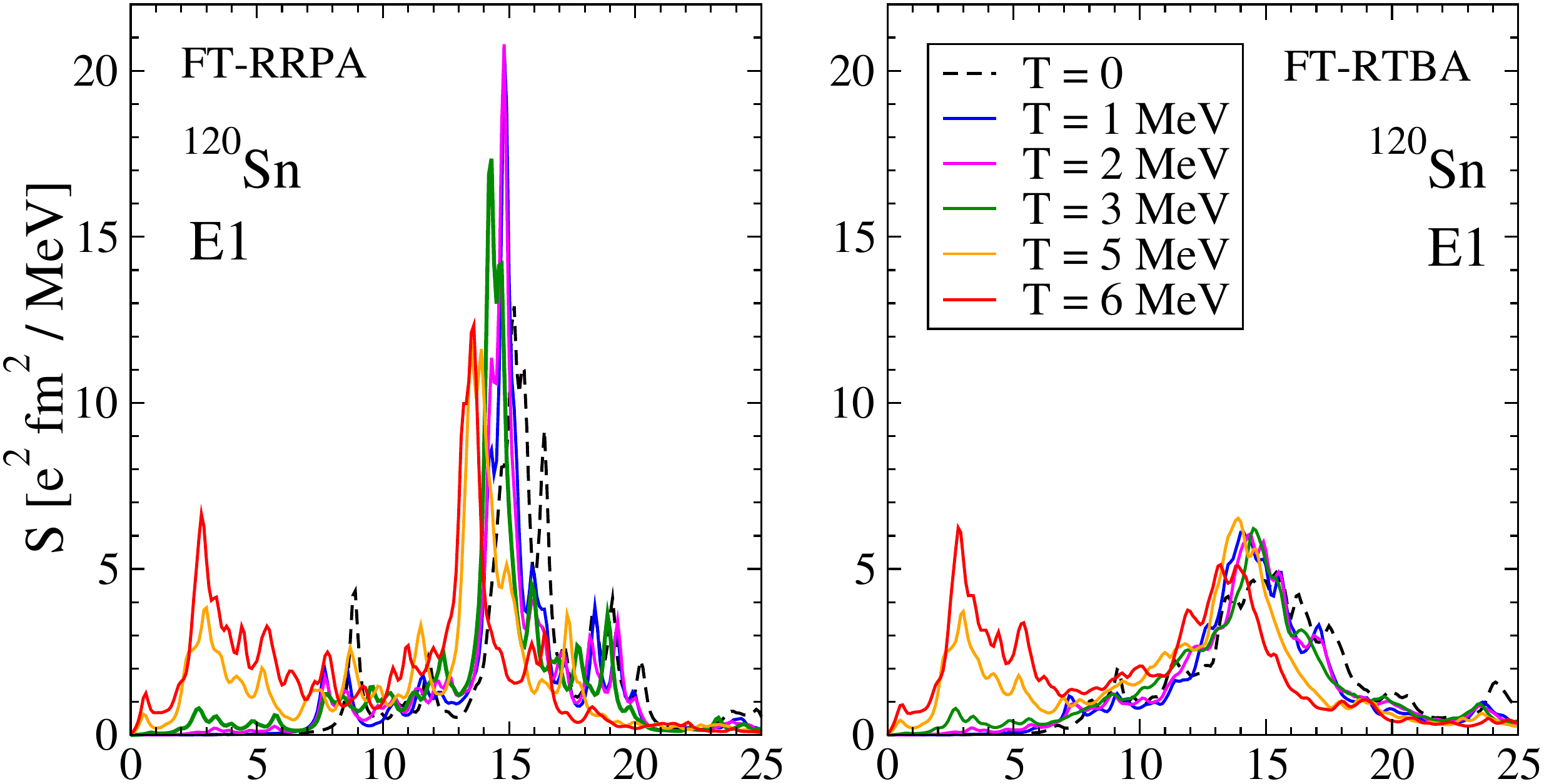}
\includegraphics[scale=0.35]{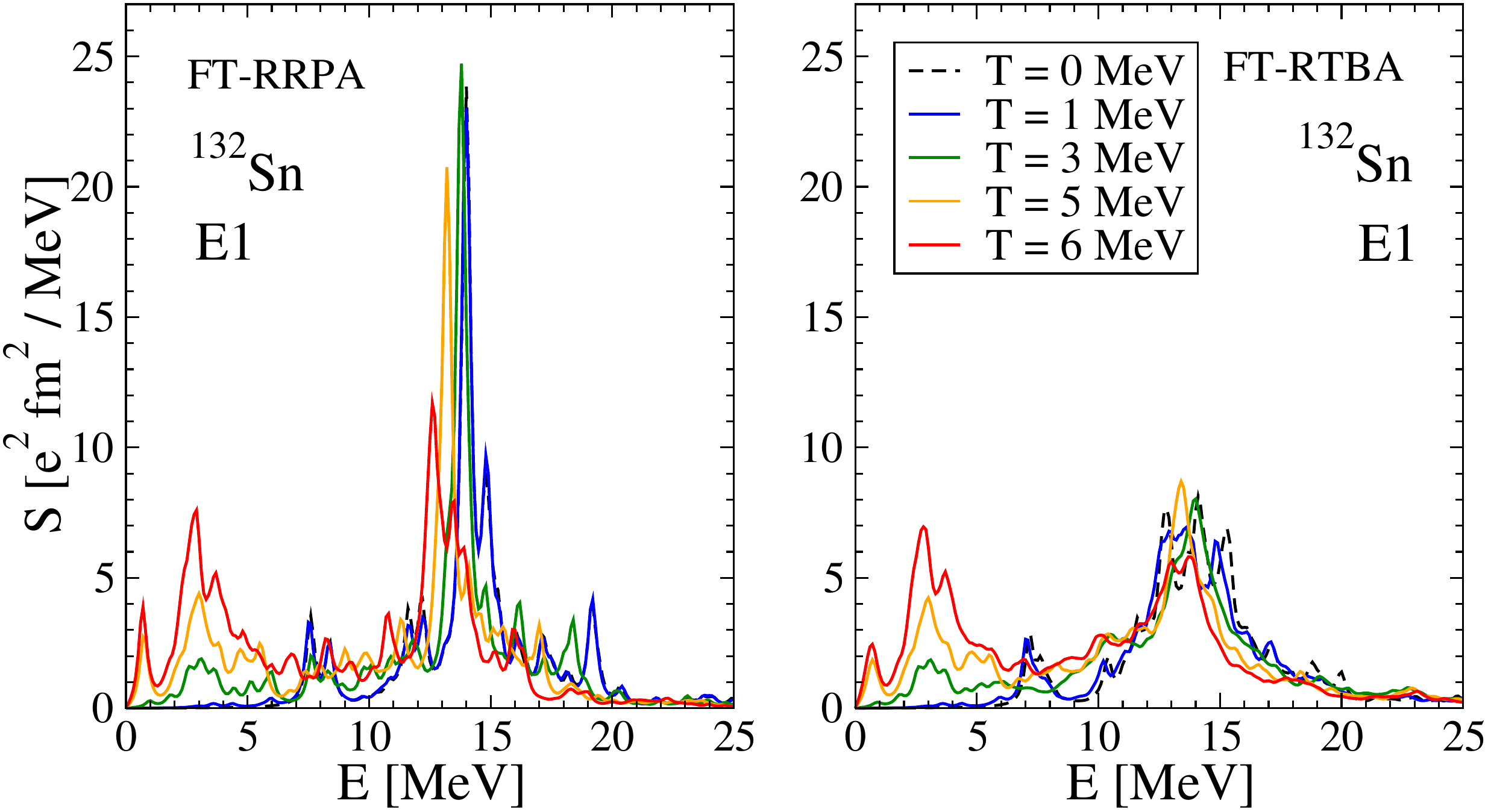}
\caption{Same as in Fig. \ref{CaNi} but for $^{100,120,132}\text{Sn}$ nuclei with the smearing parameter $\Delta=200$ keV.}
\label{Sn}
\end{figure}
\begin{figure}
\includegraphics[scale=0.45]{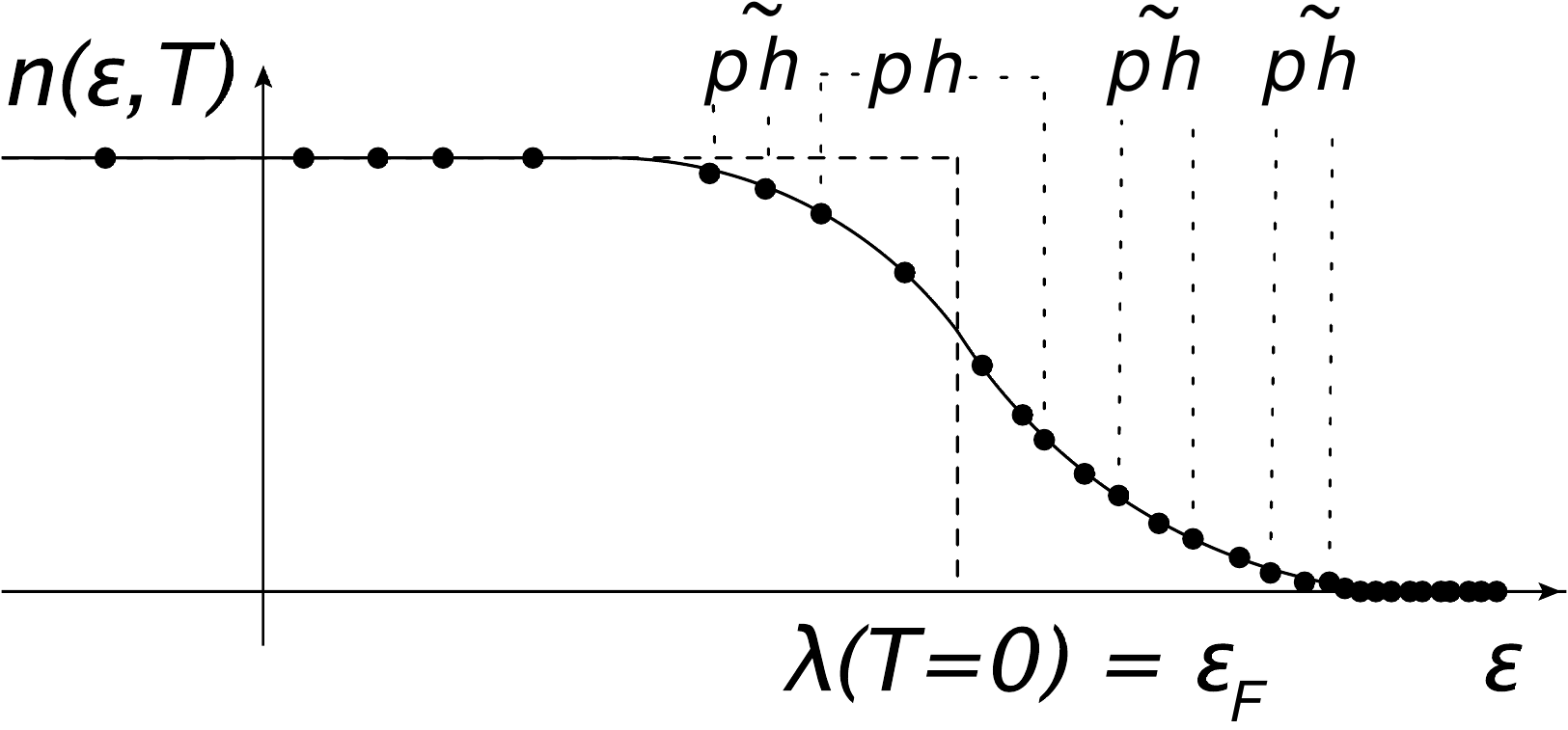}
\caption{Emergence of thermally unblocked states below and above the Fermi energy $\varepsilon_{F}$. Here $\widetilde{\textrm{ph}}$ stands for the thermally unblocked hole-hole ($\textrm{hh}$) and particle-particle ($\textrm{pp}$) fermionic pairs with the non-zero values of the uncorrelated propagator (\ref{Free Response}).}
\label{Thermally Unblocked States}
\end{figure}

The calculated temperature-dependent spectral densities $S(\omega)$ for $^{48}$Ca, $^{68}\text{Ni}$ and $^{100,120,132}\text{Sn}$ nuclei at various temperatures are shown in Figs. \ref{CaNi} and \ref{Sn}, respectively, where we compare the evolution of the electric dipole spectral density within FT-RRPA (left panels) and FT-RTBA (right panels). As the temperature increases, we observe the following two major effects: 
\begin{enumerate}
\item[(i)] The fragmentation of the dipole spectral density becomes stronger, so that the GDR undergoes a continuous broadening.  The increased diffuseness of the Fermi surface enhances significantly the amount of thermally unblocked states, especially the ones above the Fermi energy $\varepsilon_{F}$, as shown schematically in Fig. \ref{Thermally Unblocked States}. These states give rise to the new transitions within the thermal particle-hole pairs $\widetilde{\textrm{ph}}$, as follows from the form of the uncorrelated propagator (\ref{Free Response}). The increasing amount of these new pairs reinforces the Landau damping of the GDR. The spreading width of the GDR determined by the PVC amplitude of Eq. (\ref{Particle-Vibration Coupling Amplitude}) also increases because of the increasing role of the new terms with $\eta_{m}=-1$, in addition to the terms with $\eta_{m}=+1$ which solely define the PVC at zero temperature. As these terms are associated with the new poles, they enhance the spreading effects with the temperature growth, in addition to the reinforced Landau damping. At high temperatures $T\approx 5-6$ MeV, when the low-energy phonons develop the new sort of collectivity, the coupling vertices increase accordingly, which leads to a reinforcement of the spreading width of the GDR. This is consistent with the experimental observations of the "disappearance" of the high-frequency GDR at temperatures $T\geq 6$ MeV reported in the Ref. \cite{Santonocito2006}, while these temperatures might be at the limits of existence of the considered atomic nuclei.
 
\item[(ii)] The formation and enhancement of the low-energy strength below the pygmy dipole resonance. This enhancement occurs due to the new transitions within thermal $\widetilde{\textrm{ph}}$ pairs with small energy differences. The number of these pairs increases with the temperature growth in such a way that at high temperature $T\approx$ 5-6 MeV the formation of new collective low-energy modes becomes possible. Within our model, these new low-energy modes are not strongly affected by PVC. The lack of fragmentation is due to the fact that for the thermal $\widetilde{\textrm{ph}}$ pairs with small energy differences the numerator of Eq. (\ref{Particle-Vibration Coupling Amplitude}) contains the factors $n(\varepsilon_{k_{6}}-\eta_{m}\omega_{m},T)-n(\varepsilon_{k_{5}},T)$ which are considerably smaller than those for the regular $T=0$ $\textrm{ph}$ pairs of states located on the different sides with respect to the Fermi surface. Notice that the smallness of this factor for the $\widetilde{\textrm{ph}}$ pairs is not compensated by the denominator $n(\varepsilon_{k_{4}},T)-n(\varepsilon_{k_{3}},T)$ which is balanced by the numerator of Eq. (\ref{Free Response}). The inclusion of the finite-temperature ground state correlations (GSC) induced by the PVC in the particle-phonon coupling amplitude ${\Phi}(\omega)$ may enforce the fragmentation of the low-energy peak.   
\end{enumerate}

The trends are similar for the dipole strength in all considered nuclei shown in Figs. \ref{CaNi} and \ref{Sn}. The open-shell nuclei, such as $^{68}$Ni and $^{120}$Sn, are superfluid below the critical temperature which is $T_c \approx 0.6 \Delta_c$, where $\Delta_c$  is the superfluid pairing gap. It takes the values $\Delta_c = 1.6$ MeV and $\Delta_c = 1.1$ MeV for $^{68}$Ni and $^{120}$Sn, respectively, so that the superfluidity already vanishes at $T = 1$ MeV in these nuclei. As our approach does not take the superfluid pairing into account at $T>0$, we can not track this effect continuously, however, by comparing the strength distributions at $T=0$ and  $T = 1$ MeV for $^{68}$Ni and $^{120}$Sn we can see how the disappearance of superfluidity influences the strength. In the doubly-magic nuclei the dipole strength shows almost no change when going from $T=0$ to $T = 1$ MeV. This observation is consistent with the thermal RMF calculations displayed in Fig. \ref{Level Density Parameter}. As already discussed above, the presence of the large shell gaps in both neutron and proton subsystems requires a certain value of temperature to promote the nucleons over the shell gap. One can see that this temperature is $T\approx$ 0.75-1 MeV for the considered closed-shell nuclei. 

\begin{figure}[ptb]
\begin{center}
\includegraphics[scale=0.37]{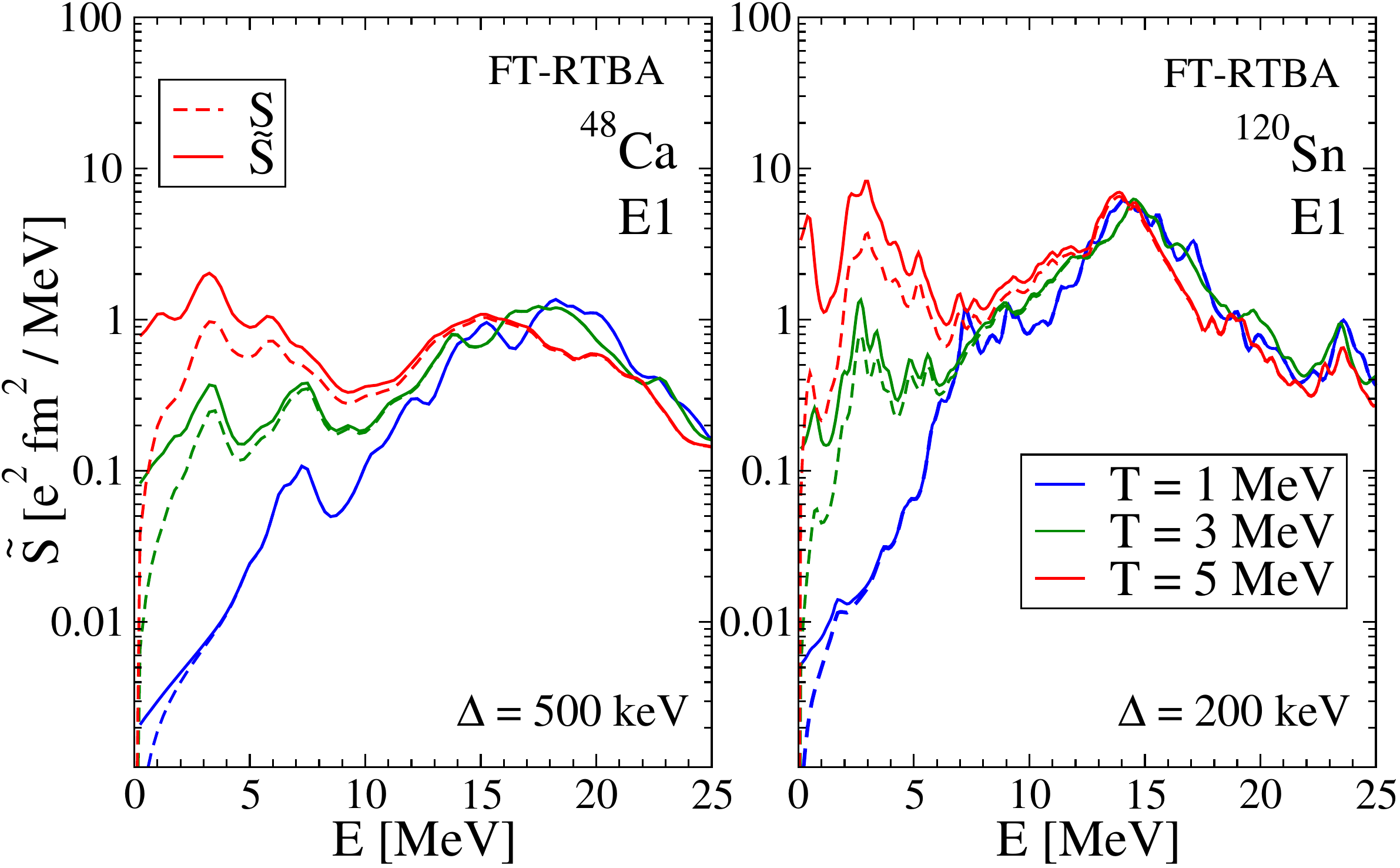}
\end{center}
\vspace{-0.5cm}
\caption{The role of the exponential factor: strength function ${\tilde S}(E)$ (solid curves) versus spectral density ${S}(E)$ (dashed curves) for the dipole strength in $^{48}$Ca (left panel) and $^{120}$Sn (right panel).}
\label{Exp_factor}
\end{figure}
%

Notice that until now  we discussed the microscopic spectral density ${S}(E)$ without the exponential factor $1/(1-$exp$(-E/T))$, which is present in the strength function ${\tilde{S}}(E)$ (\ref{Strength}) due to the detailed balance. This factor does not affect the GDR region at all temperatures under study, however, at moderate to high temperatures it enhances noticeably the low-energy strength, as illustrated in Fig. \ref{Exp_factor} for the dipole response of $^{48}\text{Ca}$ and $^{120}\text{Sn}$. At $E=0$ this factor is singular while the spectral density is equal to zero, so that the total strength function has, thus, a non-trivial limit at $E\to 0$. As one can see from Fig. \ref{Exp_factor}, this limit is finite except for the $T=0$ case when the strength function coincides with the spectral density vanishing at $E\to 0$.
In this work we focus mostly on the features of the spectral density, which is the zero-temperature analog of the strength function, in order to resolve clearly the details of the nuclear response at very low transition energies $E$ without concealing its fine features by the exponential factor. It can be easily included, for instance, when experimental data on the low-energy response become available. The important features are, in particular, the absence of the spurious translational mode and the clear zero-energy limit of the spectral density.   

\begin{figure}[ptb]
\begin{center}
\includegraphics[scale=0.37]{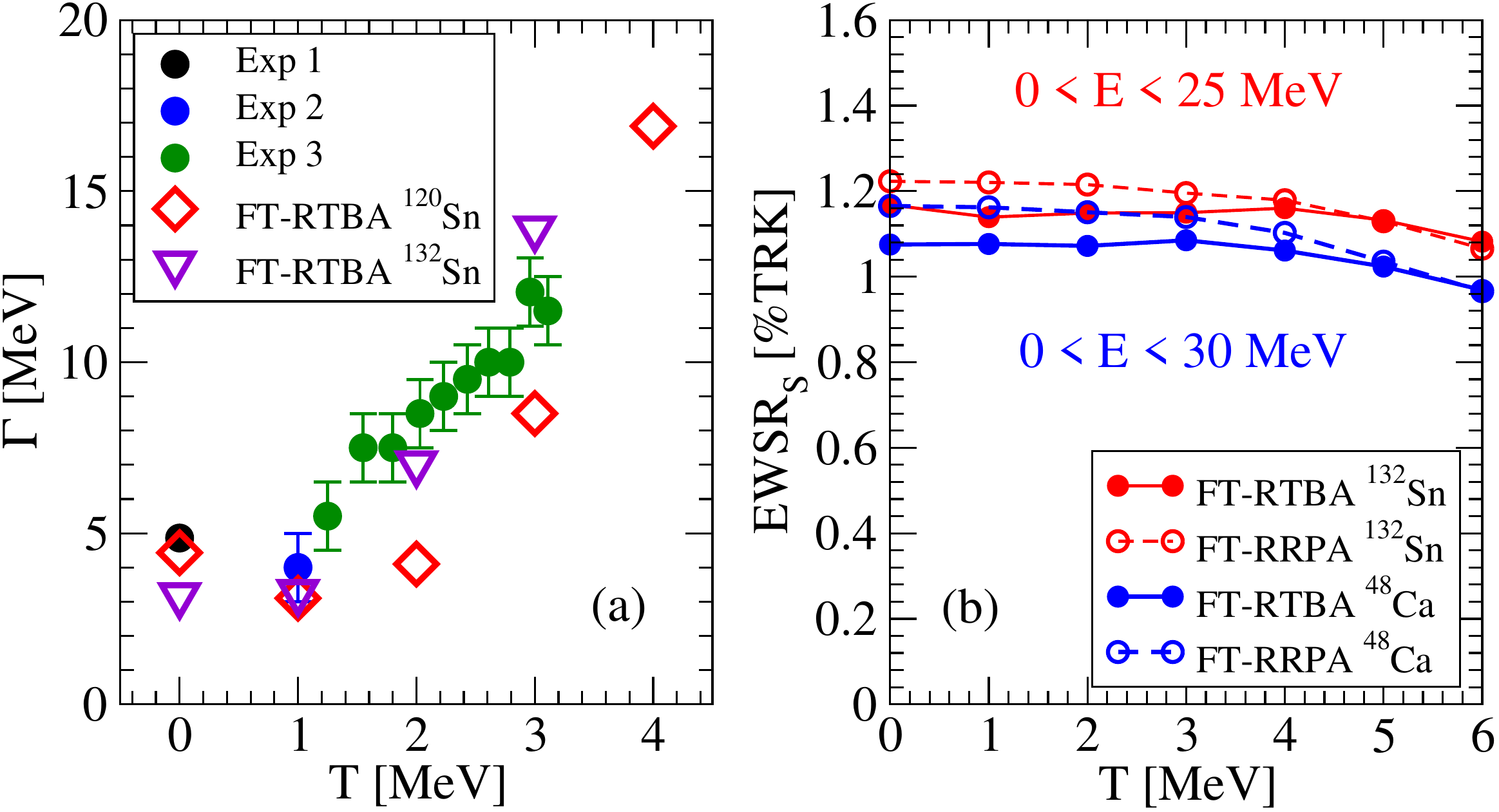}
\end{center}
\vspace{-0.5cm}
\caption{Left panel: Width of the giant dipole resonance in $^{120,132}$Sn as a function of temperature. The experimental values from Refs. \cite{Fultz1969,Heckman2003,Ramakrishnan1996} are shown for $^{120}$Sn. Right panel: The energy-weighted sum rule (EWSR) for $^{48}$Ca and $^{132}$Sn with respect to the TRK sum rule.}
\label{Gamma_EWSR}
\end{figure}

The width and the energy weighted sum rules are the most important integral characteristics of the GDR which are usually addressed in theoretical and experimental studies. In particular, they help benchmarking the theoretical approaches because of their almost model-independent character. The left panel of Fig. \ref{Gamma_EWSR} illustrates the evolution of GDR's width $\Gamma(T)$ with  temperature obtained in FT-RTBA for $^{120}\text{Sn}$ and $^{132}\text{Sn}$ nuclei together with experimental data which are available only for $^{120}$Sn. The theoretical widths at $T=0$ are taken from our previous calculations \cite{LitvinovaRingTselyaev2008,LitvinovaRingTselyaev2007}, respectively. 
Because of the phase transition in $^{120}$Sn at $T<1$ MeV, $\Gamma(T)$ has a smaller value at $T=1$ MeV than at $T=0$ as the disappearance of the superfluid pairing reduces the width.
As already mentioned, the thermal unblocking effects do not yet appear at $T=1$ MeV in both $^{120}$Sn and $^{132}$Sn
because of their specific shell structure. For the protons which form the $Z=50$ closed shell and have the next available orbitals only in the next major shell,  $T=1$ MeV temperature is not yet sufficient to promote them over the shell gap with a noticeable occupancy. In the neutron subsystem, the situation in $^{132}$Sn is similar while in $^{120}$Sn the lowest available orbit is the intruder $1h_{11/2}$ state where particles get promoted relatively easily, but after this orbit there is another shell gap. As a consequence, at $T=1$ MeV there is still no room for the $\widetilde{\textrm{ph}}$ pair formation and, hence, for a noticeable thermal unblocking. Thus, our result can explain the unexpectedly small GDR's width at $T=1$ MeV reported in Ref. \cite{Heckman2003}, in contrast to the thermal shape fluctuation calculations. After $T=1$ MeV in $^{132}$Sn and $T=2$ MeV in $^{120}$Sn we obtain a fast increase of $\Gamma(T)$ because of the formation of the low-energy shoulder by $\widetilde{\textrm{ph}}$ pairs and due to a slow increase of the fragmentation of the high-energy peak emerging from the finite-temperature effects in the PVC amplitude $\Phi(\omega)$.
 As $^{132}$Sn is more neutron-rich than $^{120}$Sn, the  respective strength in the low-energy shoulder of $^{132}$Sn is larger, which leads to a larger overall width in $^{132}$Sn at temperatures above 1 MeV. The GDR's widths for $T>3$ MeV in $^{132}$Sn and for $T>4$ MeV in $^{120}$Sn are not presented because the standard procedure based on the Lorentzian fit of the microscopic strength distribution fails in recognizing the distribution as a single peak structure.

The overall agreement of FT-RTBA calculations with data for the GDR's width in $^{120}$Sn is found very reasonable except for the temperatures around 2 MeV, possibly due to  deformation and shape fluctuation effects, which are not included in the present calculations. Our results are also consistent with those of microscopic approach of Ref. \cite{Bortignon1986}, which are available for the GDR energy region at $T\leq 3$ MeV,
while in the entire range of temperatures under study $\Gamma_{GDR}(T)$ shows a nearly quadratic dependence agreeing with the Fermi liquid theory \cite{Landau1957}. Table \ref{table} shows a comparison of $\Gamma_{GDR}(T)$ in $^{120}$Sn calculated within FT-RRPA and FT-RTBA by fitting the corresponding strength distribution by the Lorentzian  within the energy interval $0\leq E \leq 25$ MeV. One can see that in both approaches, after passing the minimum at $T=1$ MeV because of the  transition to the non-superfluid phase, $\Gamma_{GDR}(T)$ grows quickly with temperature. The difference between the width computed in the two models is about 1.0-1.7 MeV at low temperatures while it increases to $\approx 2.5$ MeV at $T=4$ MeV. It can be concluded that the PVC contribution to the width evolution is rather minor and the latter occurs mostly due to the reinforcement of the Landau damping with the temperature growth. Indeed, we could observe from varying the boundaries of the energy interval, where the fitting procedure is performed, that the amount of the low-energy strength is very important for the value of the width.

\setlength{\tabcolsep}{8pt} 
\renewcommand{\arraystretch}{1.5} 

\begin{table}[ptb]
\caption {Widths of the giant dipole resonance in $^{120}$Sn calculated by fitting the FT-RRPA and FT-RTBA strengths with the Lorentz distribution within the energy interval $0\leq E \leq 25$ MeV.} \label{table}
\begin{tabular}{llllll}
\hline
\hline
T [MeV] &  0&  1.0&  2.0&  3.0& 4.0\\
\hline
$\Gamma$ [MeV], FT-RRPA &  2.70&  2.26& 3.09 & 6.94 & 14.46\\
$\Gamma$ [MeV], FT-RTBA &  4.43&  3.08&  4.07&  8.46& 16.92\\
\hline
\hline
\end{tabular}
\end{table}

The right panel of Fig. \ref{Gamma_EWSR} shows the evolution of the energy-weighted sum rule for $^{48}\text{Ca}$  and $^{132}\text{Sn}$ nuclei calculated within FT-RRPA and FT-RTBA in the percentage with respect to the Thomas-Reiche-Kuhn (TRK) sum rule. The EWSR at $T>0$ can be calculated in full analogy with the case of $T=0$ \cite{Sommermann1983,Barranco1985a}. In our approach, where the meson-exchange interaction is velocity-dependent, already in RRPA and RQRPA at $T=0$ we observe up to  40\% enhancement of the 
TRK sum rule within the energy regions which are typically studied in experiments \cite{LitvinovaRingTselyaev2007,LitvinovaRingTselyaev2008}, in agreement with data. In the resonant time blocking approximation without the GSC of the PVC type the EWSR should have exactly the same value as in RPA \cite{Tselyaev2007} with a little violation when the subtraction procedure is performed \cite{Tselyaev2007,LitvinovaTselyaev2007}. Typically, at $T=0$ in the subtraction-corrected RTBA we find a few percent less EWSR in  finite energy intervals below 25-30 MeV than in RRPA, but this difference decreases if we take larger intervals. This is due to the fact that 
 in RTBA the strength distributions are more spread and if cut, leaves more strength outside the finite interval. 
A similar situation takes place at $T>0$. Fig. \ref{Gamma_EWSR} (b) shows that  the $\text{EWSR}$ decreases slowly with the temperature growth because the entire strength distribution moves down in energy. In both nuclei, the FT-RRPA and FT-RTBA EWSR values practically meet at $T=6$ MeV when their high-energy tails become less important.

\begin{figure} 
\includegraphics[scale=0.35]{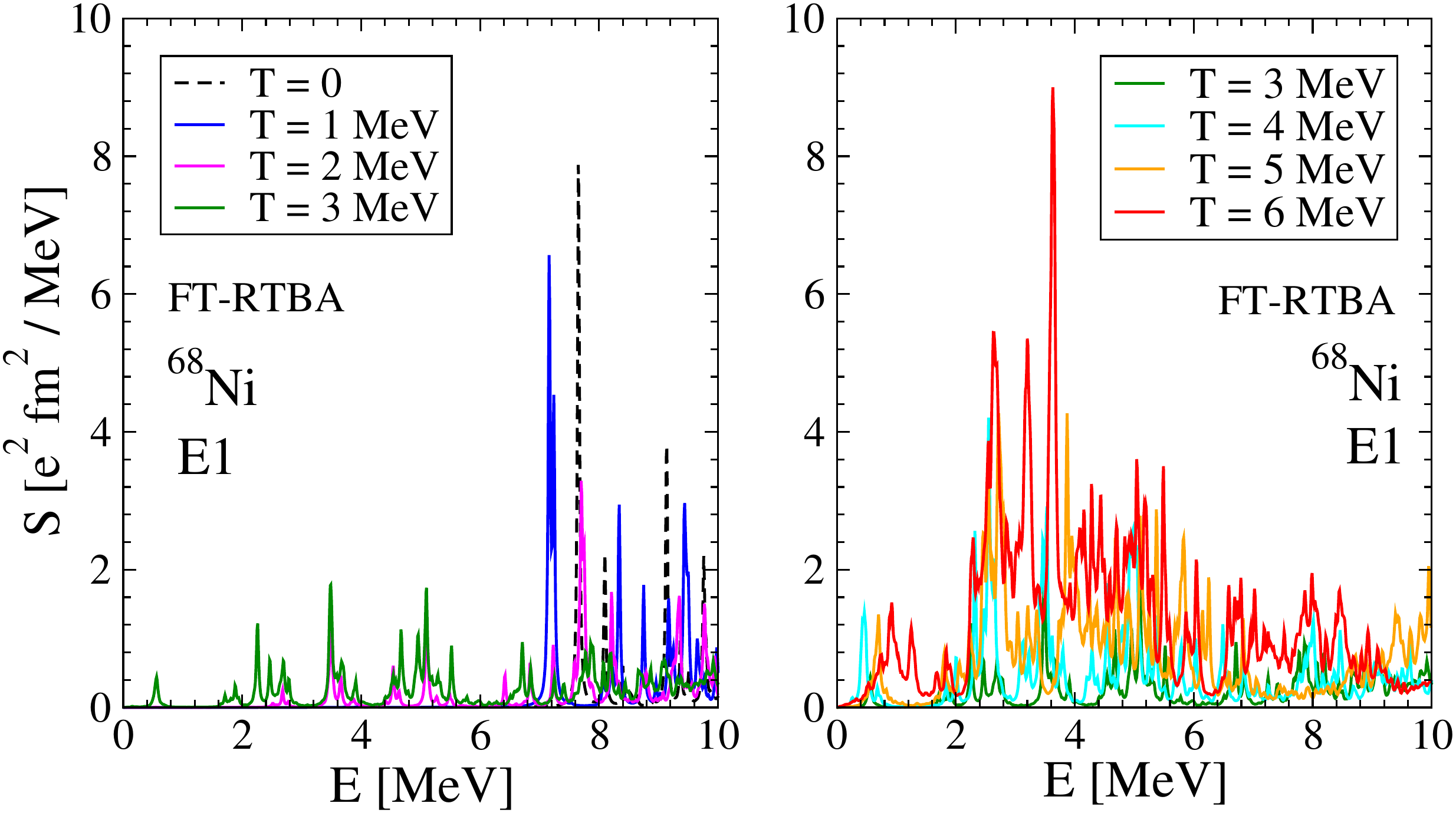}
\caption{The temperature evolution of the low-energy dipole spectral density in $^{68}$Ni calculated within FT-RTBA with the smearing parameter $\Delta = 20$ keV.}
\label{pdr_68Ni}
\end{figure}

To gain a better understanding of the formation and enhancement of the low-energy strength, we have performed a more detailed investigation of the dipole strength in the energy region $E < 10$ MeV. The dipole strength in $^{68}$Ni calculated at different temperatures with a small value of the smearing parameter $\Delta = 20$ keV is displayed in Fig. \ref{pdr_68Ni}. In the testing phase, these calculations were used to ensure positive definiteness of the spectral density as it reflects a very delicate balance between the self-energy and exchange terms in the PVC amplitude $\Phi(\omega)$. In particular, we found that consistency between $\widetilde{\textrm{ph}}$ pairs involved in self-energy and exchange terms is very important. 

%
\begin{figure*} 
\hspace{-5mm}
\includegraphics[scale=0.6]{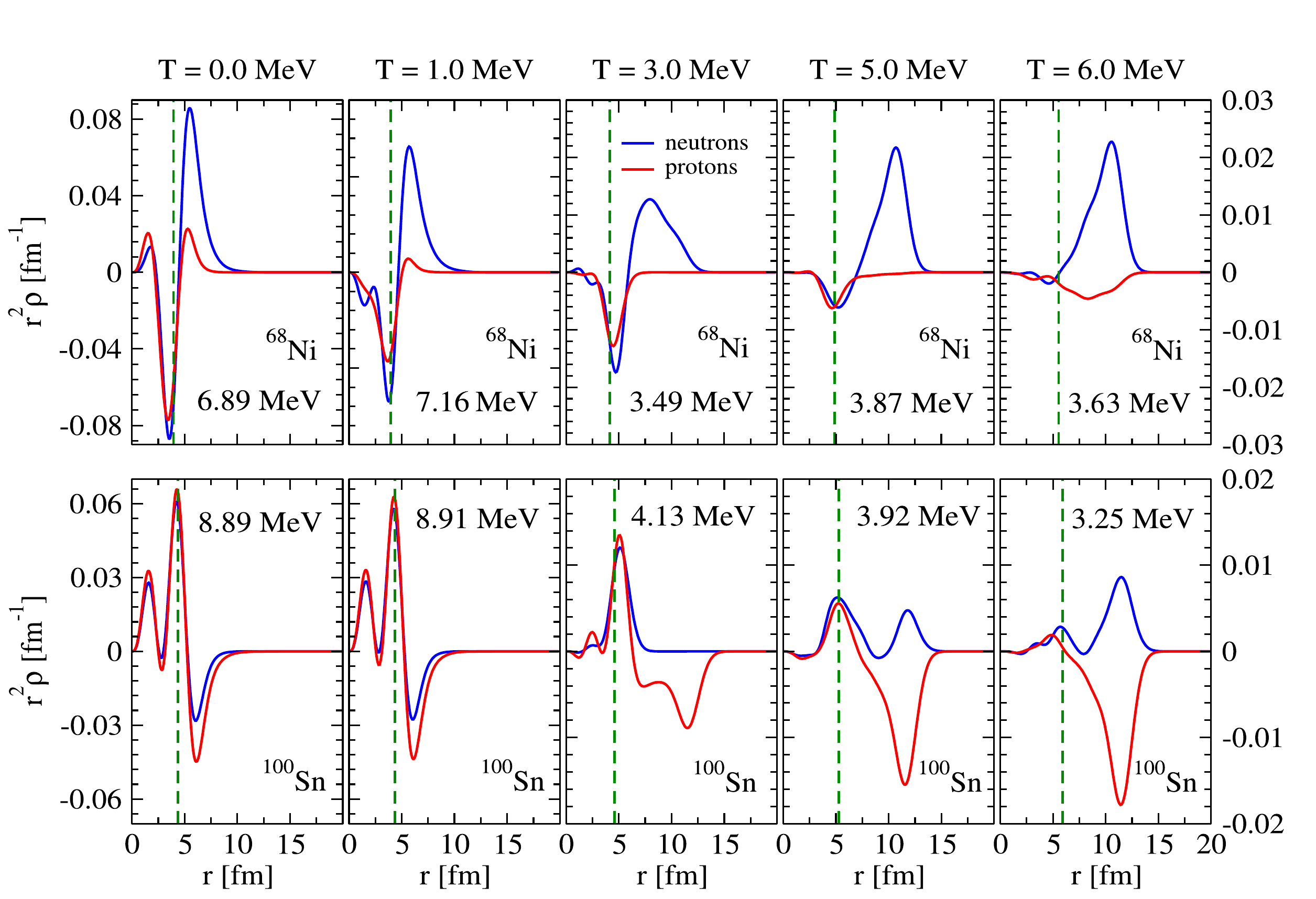}
\caption{The evolution of the proton and neutron transition densities for the most prominent peaks below 10 MeV in $^{68}\text{Ni}$ and $^{100}\text{Sn}$ within FT-RTBA. The green dashed lines indicate the rms nuclear radius.}
\label{Transition Density}
\end{figure*}
%
\begin{table*}
\caption{Major contributions of neutron (\textit{n}) and proton (\textit{p}) $\textrm{ph}$ and $\widetilde{\textrm{ph}}$ configurations to the strongest dipole states below 10 MeV in $^{68}\text{Ni}$ calculated within FT-RTBA for different temperatures.}
\label{p-n configurations}
\begin{ruledtabular}
\begin{tabular}{rlrlrlrl}
\multicolumn{2}{ c }{$^{68}\text{Ni}$; $T=0$; $E=6.89$ MeV} & \multicolumn{2}{ c }{$^{68}\text{Ni}$; $T=1$ MeV; $E=7.16$ MeV} & \multicolumn{2}{ c }{$^{68}\text{Ni}$; $T=2$ MeV; $E=7.70$ MeV} & \multicolumn{2}{ c }{$^{68}\text{Ni}$; $T=3$ MeV; $E=3.49$ MeV} \\
\hline 
10.3\% & ($2p_{3/2}\rightarrow 2d_{5/2}$) \textit{n} & 56.8\% & ($2p_{1/2}\rightarrow 3s_{1/2}$) \textit{n} & 4.9\% & ($1f_{5/2}\rightarrow 2d_{5/2}$) \textit{n} & 31.1\% & ($3s_{1/2}\rightarrow 3p_{3/2}$) \textit{n} \\
9.8\% & ($2s_{1/2}\rightarrow 2p_{3/2}$) \textit{p} & 4.4\% & ($1f_{7/2}\rightarrow 1g_{9/2}$) \textit{n} & 3.2\% & ($1f_{7/2}\rightarrow 1g_{9/2}$) \textit{n} & 15.7\% & ($2d_{5/2}\rightarrow 3p_{3/2}$) \textit{n}\\
7.1\% & ($1f_{7/2}\rightarrow 1g_{9/2}$) \textit{p} & 2.2\% & ($1f_{5/2}\rightarrow 2d_{5/2}$) \textit{n} & 2.9\% & ($2p_{3/2}\rightarrow 2d_{5/2}$) \textit{n} & 0.1\% & ($3s_{1/2}\rightarrow 3p_{1/2}$) \textit{n}\\
6.2\% & ($1f_{5/2}\rightarrow 2d_{5/2}$) \textit{n} & 1.4\% & ($1f_{7/2}\rightarrow 1g_{9/2}$) \textit{p} & 2.1\% & ($1f_{5/2}\rightarrow 2d_{3/2}$) \textit{n} & 0.01\% & ($1f_{7/2}\rightarrow 1g_{9/2}$) \textit{n}\\
6.1\% & ($1f_{7/2}\rightarrow 1g_{9/2}$) \textit{n} & 1.0\% & ($1f_{5/2}\rightarrow 2d_{3/2}$) \textit{n} & 1.7\% & ($1f_{7/2}\rightarrow 1g_{9/2}$) \textit{p} & 0.01\% & ($1g_{9/2}\rightarrow 1h_{11/2}$) \textit{n}\\
4.6\% & ($1f_{5/2}\rightarrow 2d_{3/2}$) \textit{n} & 0.9\% & ($2p_{3/2}\rightarrow 3s_{1/2}$) \textit{n} & 1.3\% & ($2p_{1/2}\rightarrow 2d_{3/2}$) \textit{n} & \\
1.0\% & ($2p_{1/2}\rightarrow 2d_{3/2}$) \textit{n} & 0.9\% & ($2p_{1/2}\rightarrow 2d_{3/2}$) \textit{n} & 1.1\% & ($2s_{1/2}\rightarrow 2p_{3/2}$) \textit{p} & \\
0.9\% & ($1d_{3/2}\rightarrow 2p_{1/2}$) \textit{p} & 0.7\% & ($2p_{1/2}\rightarrow 4s_{1/2}$) \textit{n} & 0.9\% & ($2p_{3/2}\rightarrow 3s_{1/2}$) \textit{n} & \\
0.9\% & ($1d_{3/2}\rightarrow 2p_{3/2}$) \textit{p} & 0.5\% & ($2p_{3/2}\rightarrow 2d_{5/2}$) \textit{n} & 0.2\% & ($1d_{3/2}\rightarrow 2p_{1/2}$) \textit{p} & \\
0.7\% & ($2p_{3/2}\rightarrow 3s_{1/2}$) \textit{n} & 0.3\% & ($1d_{3/2}\rightarrow 2p_{3/2}$) \textit{p} & 0.2\% & ($1d_{3/2}\rightarrow 2p_{3/2}$) \textit{p} & \\
0.4\% & ($1f_{5/2}\rightarrow 3d_{3/2}$) \textit{n} & 0.2\% & ($1d_{3/2}\rightarrow 2p_{1/2}$) \textit{p} & 0.1\% & ($1f_{5/2}\rightarrow 3d_{3/2}$) \textit{n} & \\
0.3\% & ($2s_{1/2}\rightarrow 2p_{1/2}$) \textit{p} & 0.1\% & ($2p_{1/2}\rightarrow 5s_{1/2}$) \textit{n} & \\
0.2\% & ($2p_{3/2}\rightarrow 3d_{5/2}$) \textit{n} & \\
0.2\% & ($1f_{5/2}\rightarrow 3d_{5/2}$) \textit{n} & \\
0.2\% & ($2p_{3/2}\rightarrow 2d_{3/2}$) \textit{n} & \\
0.2\% & ($1f_{7/2}\rightarrow 2d_{5/2}$) \textit{p} & \\
0.1\% & ($1f_{7/2}\rightarrow 2d_{5/2}$) \textit{n} & \\
 & & & & & & \\
49.2\% & & 69.4\% & & 18.6\% & & 46.92\% \\
\hline
\hline
\\
\hline
\hline
\multicolumn{2}{ c }{$^{68}\text{Ni}$; $T=4$ MeV; $E=2.55$ MeV} & \multicolumn{2}{ c }{$^{68}\text{Ni}$; $T=5$ MeV; $E=3.87$ MeV} & \multicolumn{2}{ c }{$^{68}\text{Ni}$; $T=6$ MeV; $E=3.63$ MeV} \\
\hline
66.1\% & ($2f_{7/2}\rightarrow 2g_{9/2}$) \textit{n} & 61.9\% & ($2g_{9/2}\rightarrow 2h_{11/2}$) \textit{n} & 21.2\% & ($1i_{11/2}\rightarrow 1j_{13/2}$) \textit{n} & \\
5.1\% & ($3p_{1/2}\rightarrow 3d_{3/2}$) \textit{n} & 3.0\% & ($3f_{7/2}\rightarrow 4d_{5/2}$) \textit{n} & 9.5\% & ($2d_{5/2}\rightarrow 2f_{7/2}$) \textit{p} & \\
0.7\% & ($2f_{5/2}\rightarrow 2g_{7/2}$) \textit{n} & 0.4\% & ($2g_{7/2}\rightarrow 3f_{5/2}$) \textit{n} & 8.8\% & ($1i_{13/2}\rightarrow 1j_{15/2}$) \textit{n} & \\
0.4\% & ($2d_{3/2}\rightarrow 3p_{1/2}$) \textit{n} & 0.3\% & ($2d_{3/2}\rightarrow 2f_{5/2}$) \textit{p} & 3.2\% & ($2d_{3/2}\rightarrow 2f_{5/2}$) \textit{n} & \\
0.1\% & ($1g_{7/2}\rightarrow 2f_{5/2}$) \textit{n} & 0.2\% & ($1h_{11/2}\rightarrow 1i_{13/2}$) \textit{n} & 0.1\% & ($2g_{9/2}\rightarrow 3f_{7/2}$) \textit{n} & \\
 & & 0.1\% & ($3d_{3/2}\rightarrow 2f_{5/2}$) \textit{n} & & & \\
 & & & & & & \\
72.4\% & & 65.9\% & & 42.8\% & & \\
\end{tabular}
\end{ruledtabular}
\end{table*}
The FT-RTBA calculations presented in Fig. \ref{pdr_68Ni} resolve individual states in the low-energy region showing the details of the evolution of the thermally emergent dipole strength. In particular, one can trace how the major peak moves toward lower energies and its intensity increases. The proton and neutron transition densities for the most prominent peak below 10 MeV are displayed for different temperature values in Fig. \ref{Transition Density} for the neutron-rich $^{68}$Ni nucleus and for the neutron-deficient $^{100}$Sn nucleus. 
In the neutron-rich $^{68}\text{Ni}$ nucleus proton and neutron transition densities show in-phase oscillations inside the nucleus while neutron oscillations become absolutely dominant outside for $0\leq T \leq 5$ MeV. At the temperature $T=6$ MeV protons and neutrons exhibit out of phase oscillation which resembles the well-recognized pattern of the collective giant resonance. Indeed, as it is shown in Table \ref{p-n configurations} below, the low-energy peak at $T=6$ MeV has some features of collective nature. 
The situation is quite similar in the neutron-deficient $^{100}\text{Sn}$ nucleus, which exhibits the in-phase oscillations of protons and neutrons inside the nucleus, but with the dominance of proton oscillations in the outer area. Analogously, at $T = 6$ MeV one starts to distinguish a GDR-like pattern of the out-of-phase oscillation in the low-lying state at $E = 3.25$ MeV. We also notice that  at $3\leq T \leq 6$ MeV the oscillations extend to far distances from the nuclear central region.

In order to have some more insights into the structure of the new low-energy states, we have extracted the $\widetilde{\textrm{ph}}$ compositions of the strongest low-energy states at various temperatures. The quantities
\begin{equation}
z_{\textrm{ph}}^{fi} = \frac{|\rho^{fi}_{\textrm{ph}}|^{2}-|\rho^{fi}_{\textrm{hp}}|^{2}}{n_{\textrm{h}}(\varepsilon_{\textrm{h}},T)-n_{\textrm{p}}(\varepsilon_{\textrm{p}},T)}
\end{equation}
are given in Table \ref{p-n configurations} in percentage with respect to the FT-RTBA generalized normalization condition of Eq. (\ref{Generalized Normalization Condition}). In most of the cases, we omit contributions of less than 0.1 \%. The bottom line shows the total percentage of pure $\textrm{ph}$ and $\widetilde{\textrm{ph}}$ configurations, so that the deviation of this number from 100 \% characterizes the degree of PVC, according to Eq. (\ref{Generalized Normalization Condition}).  

We start with the state at $E = 6.89$ MeV at $T=0$ which shows up as a slightly neutron-dominant state with seven two-quasiparticle contributions bigger than 1 \%. This state can be classified as a relatively collective one. At $T = 1$ MeV the $^{68}\text{Ni}$ nucleus becomes non-superfluid and one can see that the strongest low-energy state has a dominant particle-hole configuration. For the $E = 7.70$ MeV state at $T = 2$ MeV the major contribution comes from the PVC as the particle-hole configurations sum up to 18.6 \% only. It is important to emphasize that the considered peaks at $T\leq 2$ MeV are dominated by the $\textrm{ph}$ transitions of nucleons  across the Fermi surface, while at $T\geq 3$ MeV they are mainly composed of the thermal $\widetilde{\textrm{ph}}$ transitions between the states above the Fermi energy. These states are mostly located in the continuum, which is discretized in the present calculations. Although a more accurate continuum treatment is necessary to investigate the low-energy response at finite temperatures \cite{LitvinovaBelov2013}, as the large number of the basis harmonic oscillator shells are taken into account in this work, the discretized description of the continuum should be quite adequate. Thus, we notice that at $T\geq 2$ MeV the collectivity becomes destroyed by the thermal effects until it reappears again at $T=6.0$ MeV.  This temperature is, however, rather high and can be close to the limiting temperature which terminates existence of the nucleus \cite{Santonocito2006}. 

\section{CONCLUSIONS AND OUTLOOK}

We present a finite-temperature extension of the nuclear response theory beyond the relativistic RPA. In order to calculate the time-dependent part of the nucleon-nucleon interaction, which contains coupling between nucleons and correlated two-nucleon pairs (phonons), we generalize the time blocking method developed previously for the zero-temperature case. The proposed soft blocking applied to the Matsubara two-fermion propagators allows for ordering the corresponding diagrams in the imaginary-time domain and, thus, reduces the Bethe-Salpeter equation for the nuclear response to a single frequency variable equation.

The method named finite-temperature relativistic time blocking approximation was implemented on the base of quantum hadrodynamics which was thereby extended beyond the one-loop approximation for finite temperatures. Using the NL3 parametrization for the covariant energy density functional, we investigated the temperature dependence of the dipole response in 
medium-light $^{48}$Ca, $^{68}\text{Ni}$ and medium-heavy $^{100,120,132}\text{Sn}$ nuclei.  The obtained results are consistent with the existing experimental data on the GDR's width and with the result of Landau theory for the temperature dependence of the GDR's width.  The calculations extended to high temperatures explain the critical phenomenon of the disappearance of the GDR and suggest that the collective motion may reappear at low frequencies in the high-temperature regime. 

The analytical method presented in this work is of a general character, so that it can be widely applied to the response of strongly-correlated systems at finite temperature. 
The presented numerical implementation of FT-RTBA opens a way to quantitative systematic studies of excitations and de-excitations of compound nuclei in a wide energy range. For an accurate description of the low-energy strength at the r-process temperature conditions the present version of FT-RTBA has to be further improved by the inclusion of continuum effects and ground state correlations associated with the PVC. Future work will address these issues.

\section*{Acknowledgements}

The authors greatly appreciate discussions with Peter Schuck and Jian Li. This work is partly supported by US-NSF Grant PHY-1404343 and NSF Career Grant PHY-1654379.
%

\bibliography{Bibliography_Sep2018.bib}
\end{document}